\documentclass[fleqn,usenatbib]{mnras}

\usepackage{newtxtext,newtxmath}

\usepackage[T1]{fontenc}
\usepackage{ae,aecompl,times}

\usepackage{graphicx}	
\usepackage{amsmath}	
\usepackage{amssymb}	
\usepackage{multirow}
\usepackage[normalem]{ulem}
\usepackage{xcolor}
\usepackage{bm}
\usepackage[normalem]{ulem}

\newcommand{\Mpch}{\,h^{-1}{\rm Mpc}}
\newcommand{\hMpc}{\,h\,{\rm Mpc}^{-1}}
\newcommand{\hMpcc}{\,h^3{\rm \,Mpc}^{-3}}
\newcommand{\Msh}{\,h^{-1}{\rm M}_\odot}
\newcommand{\lla}{\left\langle}
\newcommand{\rra}{\right\rangle}
\newcommand{\Mr}{^{0.1}M_r - 5\mathrm{log}_{10} h}
\newcommand{\gr}{^{0.1}(g-r)}
\newcommand{\Mmean}{M_{200\mathrm{m}}}

\definecolor{mred}{rgb}{0.058, 0.588, 0.778}

\title[Measuring the BAO peak position]{Measuring the BAO peak position with different galaxy selections}

\author[C.~Hern\'andez-Aguayo et al.]
{
\parbox{0.99\textwidth}{C\'esar Hern\'andez-Aguayo$^{1}$\thanks{E-mail: 
cesar.hernandez-aguayo@durham.ac.uk (CH-A)}, Marius Cautun$^{2,1}$, Alex Smith$^{3,1}$, Carlton M. Baugh$^{1}$ and Baojiu Li$^{1}$}
\\
\\
$^{1}$Institute for Computational Cosmology, Department of Physics, Durham University, South Road, Durham, DH1 3LE, UK.\\
$^{2}$Leiden Observatory, Leiden University, PO Box 9513, NL-2300 RA Leiden, the Netherlands.\\
$^{3}$IRFU, CEA, Universit\'e Paris-Saclay, F-91191 Gif-sur-Yvette, France.
}

\date{Accepted XXX. Received YYY; in original form ZZZ}

\pubyear{2019}

\begin{document}
\label{firstpage}
\pagerange{\pageref{firstpage}--\pageref{lastpage}}
\maketitle

\begin{abstract}
We investigate if, for a fixed number density of targets and redshift, there is an optimal way to select a galaxy sample in order to measure the baryon acoustic oscillation (BAO) scale, which is used as a standard ruler to constrain the cosmic expansion. Using the mock galaxy catalogue built by Smith et al. in the Millennium-XXL N-body simulation with a technique to assign galaxies to dark matter haloes based on halo occupation distribution modelling, we consider the clustering of galaxies selected by luminosity, colour and local density. We assess how well the BAO scale can be extracted by fitting a template to the power spectrum measured for each sample.  
We find that the BAO peak position is recovered equally well for samples defined by luminosity or colour, while there is a bias in the BAO scale recovered for samples defined by density.
The BAO position is contracted to smaller scales for the densest galaxy quartile and expanded to large scales for the two least dense galaxy quartiles. For fixed galaxy number density, density-selected samples have higher uncertainties in the recovered BAO scale than luminosity- or colour-selected samples.
\end{abstract}




\section{Introduction}
\label{sec:intro}
The baryon acoustic oscillations (BAO) scale is a standard ruler that can be used to measure the cosmological redshift - distance relation \citep{Eisenstein:1997ik,Blake:2003rh,Linder:2003ec,Xu:2012fw,Ross:2014qpa}. This characteristic scale is approximately the horizon scale at recombination and corresponds to the largest distance that a sound wave can travel in the photon -- baryon fluid up to this epoch. 
The sound horizon at recombination has been measured at the sub-percent level using the cosmic microwave background (CMB) radiation \citep{Planck:2016}.
It is possible to measure the BAO scale from the clustering of galaxies using two-point statistics such as the correlation function or its Fourier transform, the power spectrum \citep[see e.g.,][]{Cole:2005,Eisenstein:2005su,Beutler:2016ixs,Ross:2016gvb}. This allows us to probe the redshift -- distance relation, which depends on the cosmological model and hence, given the existing constraints from the CMB, constrains the late-time behaviour of the dark energy.

A variety of tracers are being considered to probe the large-scale structure of the Universe over different redshift intervals \citep[see][]{Laureijs:2011gra,Amendola:2012ys,DESI:2016,Alam:2017hwk}. For example, the Dark Energy Spectroscopic Instrument (DESI) survey will carry out four galaxy surveys \citep{DESI:2016}: i) a magnitude limited sample at low redshifts, ii) luminous red galaxies (LRGs) at intermediate redshifts up $z \sim 1$, iii) emission line galaxies (ELGs) to $z \sim 1.7$ and iv) quasi-stellar objects (QSOs) at $z < 2.1$. Different targeting strategies are driven partly by observational and instrumental considerations, such as the visibility of a particular emission line over a given redshift interval or the number of fibres available in the field of view. Our aim here is to assess the relative merits of using different galaxy selections to  measure the BAO scale.

\begin{figure*}
    \centering
    \includegraphics[width=0.45\textwidth]{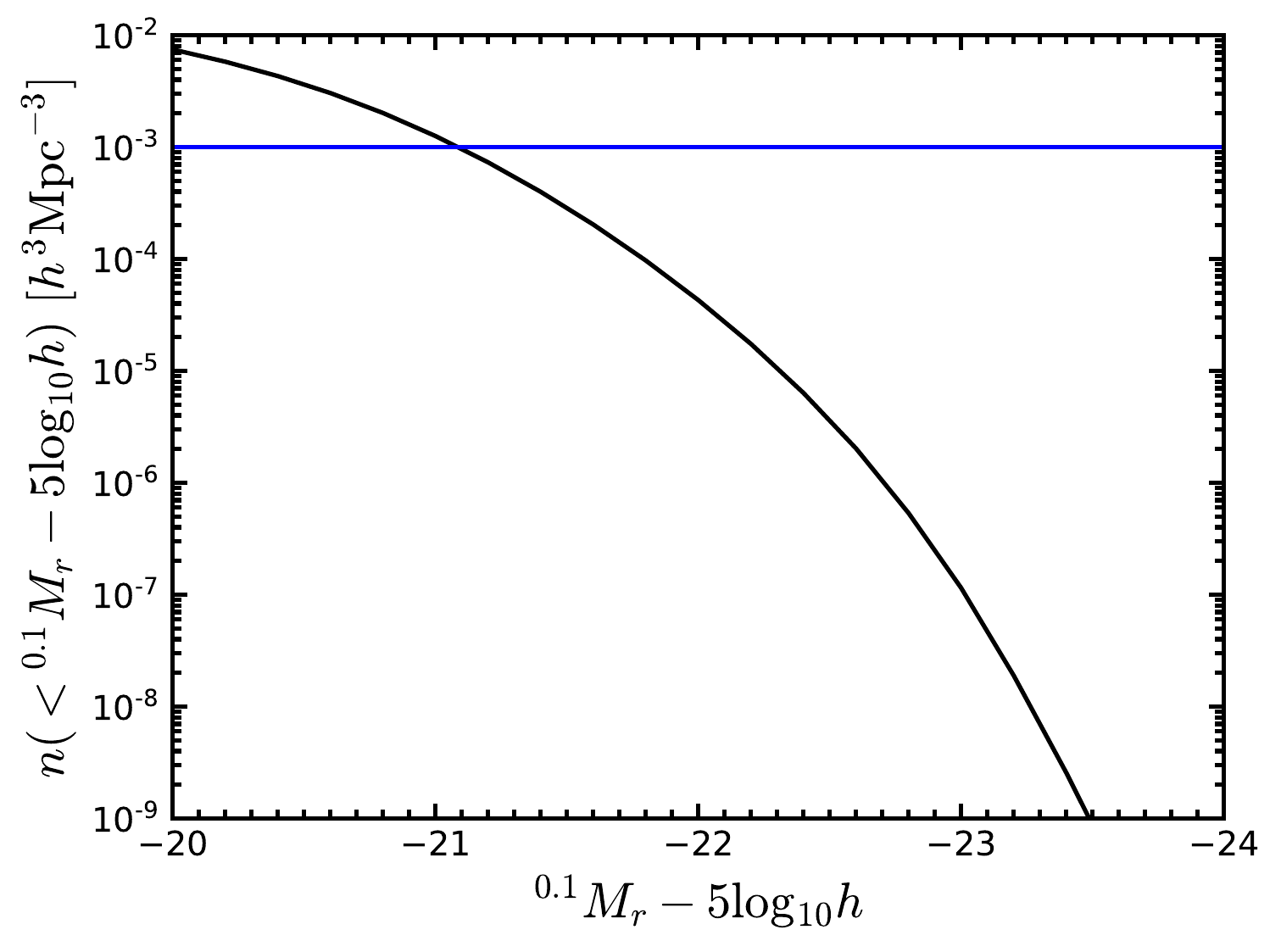}
    \includegraphics[width=0.45\textwidth]{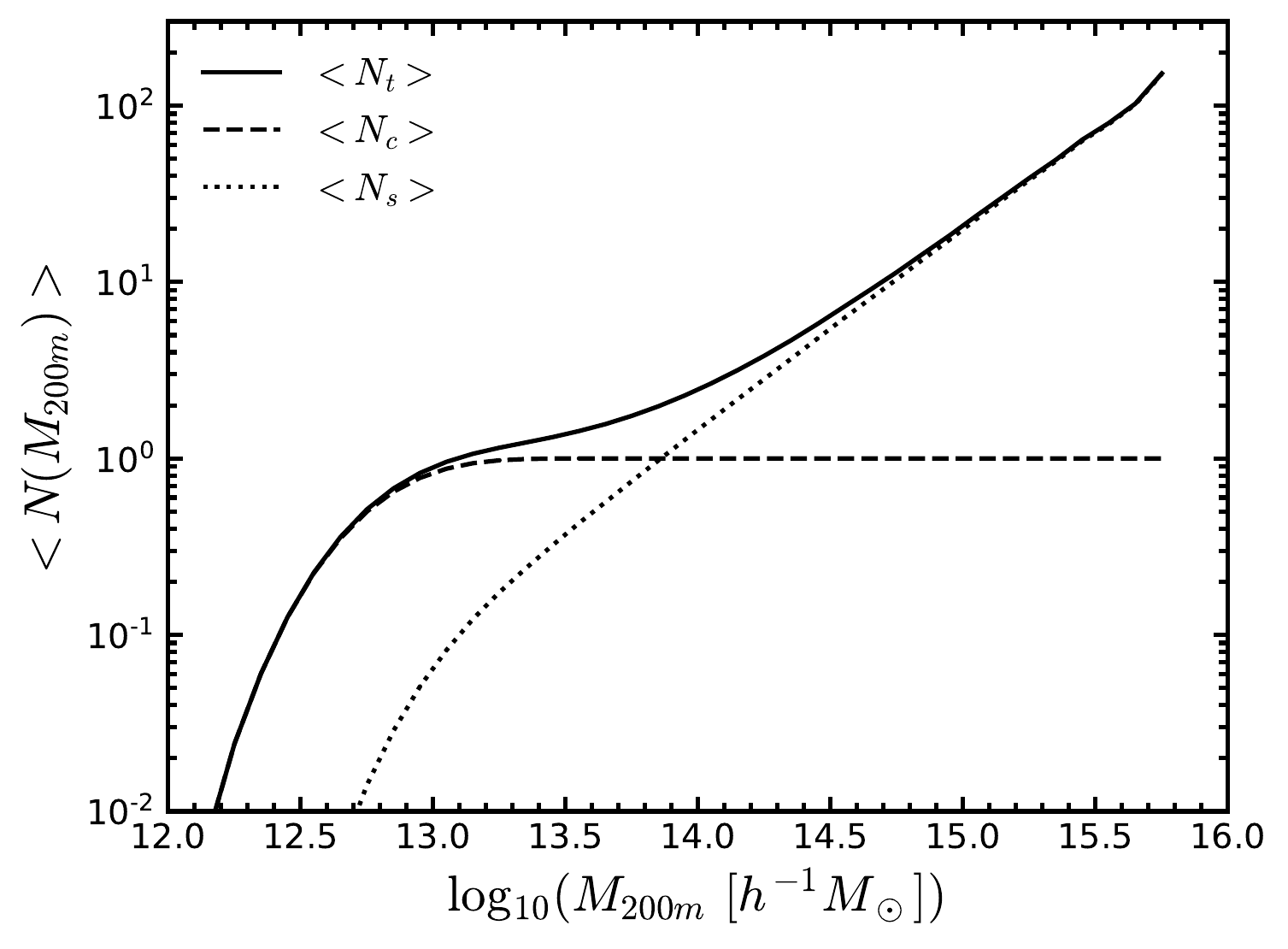}
    \caption{{\it Left panel:} $r$-band cumulative luminosity function of the parent galaxy catalogue at $z=0.11$. 
    The horizontal blue line indicates the number density of the full sample, $n=10^{-3} h^{3} \, {\rm Mpc}^{-3}$, which corresponds to retaining galaxies brighter than a magnitude cut of $^{0.1}M_r - 5\log_{10} h = -21.08$.
    {\it Right panel:} Halo occupation distribution of the full sample. The occupation functions of all, central and satellite galaxies are shown as solid, dashed and dotted lines, as specified in the legend.}
    \label{fig:cLF}
\end{figure*}

We explore if there is an optimal way to target galaxies to extract the BAO scale. We do this by ranking galaxies by either their luminosity, colour, or environment within the same volume, and then assess how well we can extract the BAO scale for different subsamples of galaxies in each case. The initial idea behind using subsamples of galaxies was to sparsely sample a flux limited catalogue to efficiently map a large survey volume, without measuring a redshift for every galaxy \citep{Kaiser:1986}. This technique was successfully applied to early redshift surveys to yield impressive constraints on cosmological parameters from modest numbers of galaxy redshifts  \citep{Esfathiou:1990,Loveday:1992}. A development of this approach was to target a particular class of object rather than to randomly sample a flux limited catalogue. LRGs were isolated from the photometric catalogue of the Sloan Digital Sky Survey to probe a larger volume of the Universe than that reached by the original flux limited survey \citep{Eisenstein:2001}. The argument here is that the LRGs should be strongly biased tracers of the underlying dark matter, because they are bright galaxies, thereby boosting the signal-to-noise of the clustering measurement for a fixed number density of targets \citep{Feldman:1994}. Similar strategies were devised to map the large-scale structure of the Universe out to $z \sim 1$ using galaxies with strong emission lines (ELGs) \citep{Drinkwater:2010}. Recently, \cite{Ruggeri:2019kjl} re-analysed the data from the 6dFGS, WiggleZ, BOSS and eBOSS galaxy surveys to study how assumptions about the errors and sample variance affect the recovery of the BAO scale. 

Characterising how the BAO signal varies between different galaxy populations is also important for understanding systematic biases in the position of the BAO peak. For example, overdense regions contract, pulling the BAO peak inwards, while underdense regions expand, pushing the BAO peak to larger scales \citep{Sherwin:2012,Neyrinck:2016pfm}. Different galaxy populations sample the underlying density field differently and thus the size of this effect can vary between galaxy populations \citep[e.g.][]{Angulo:2008,McCullagh:2013,Achitouv:2015}. Such systematic effects are small, but nonetheless are important for current and future precision measurements. To a first approximation, this effect, as well as the smearing of the BAO peak, is captured by ``BAO reconstruction" techniques, such at those based on Lagrangian linear theory \citep[e.g.][]{Eisenstein2007ApJ...664..675E,Padmanabhan2012} and the more recent non-linear reconstruction techniques \citep[e.g.][]{Ata2015,Zhu2017,Hada:2018,Shi:2018,Birkin:2019,Jasche2019}. However, these methods are rather involved and it remains to be understood if they fully account for the BAO systematics present in different galaxy samples. This is why here we study the BAO signal in the galaxy distribution without applying a BAO reconstruction step. 

To address the question of what is the best way to measure BAO, we use a mock catalogue built by implementing a technique based on halo occupation distribution modelling into one of the largest N-body simulations ever run, the Millennium-XXL \citep{Angulo:2012ep,Smith:2017tzz}. We test how well the BAO scale can be constrained for galaxy samples selected in different ways using a power spectrum analysis. 
Our goal is to establish how the strength of the BAO feature, and thus the accuracy with which the BAO scale can be measured, depends on galaxy properties such as  brightness, colour and local density. In particular, we investigate what are the best ways to select galaxies such that we optimise the BAO measurement for future spectroscopic surveys. The results of our study can inform the survey strategy of upcoming projects.

The paper is organised as follows: In Section~\ref{sec:data}, we describe the construction of the galaxy catalogue and the theoretical BAO model. In Sec.~\ref{sec:clustering}, we show the results of the power spectrum fitting and a description of the galaxy-halo connection of the galaxy samples. Finally, the summary and our conclusions are given in Sec.~\ref{sec:conc}.

\begin{figure*}
 \centering
\includegraphics[width=0.45\textwidth]{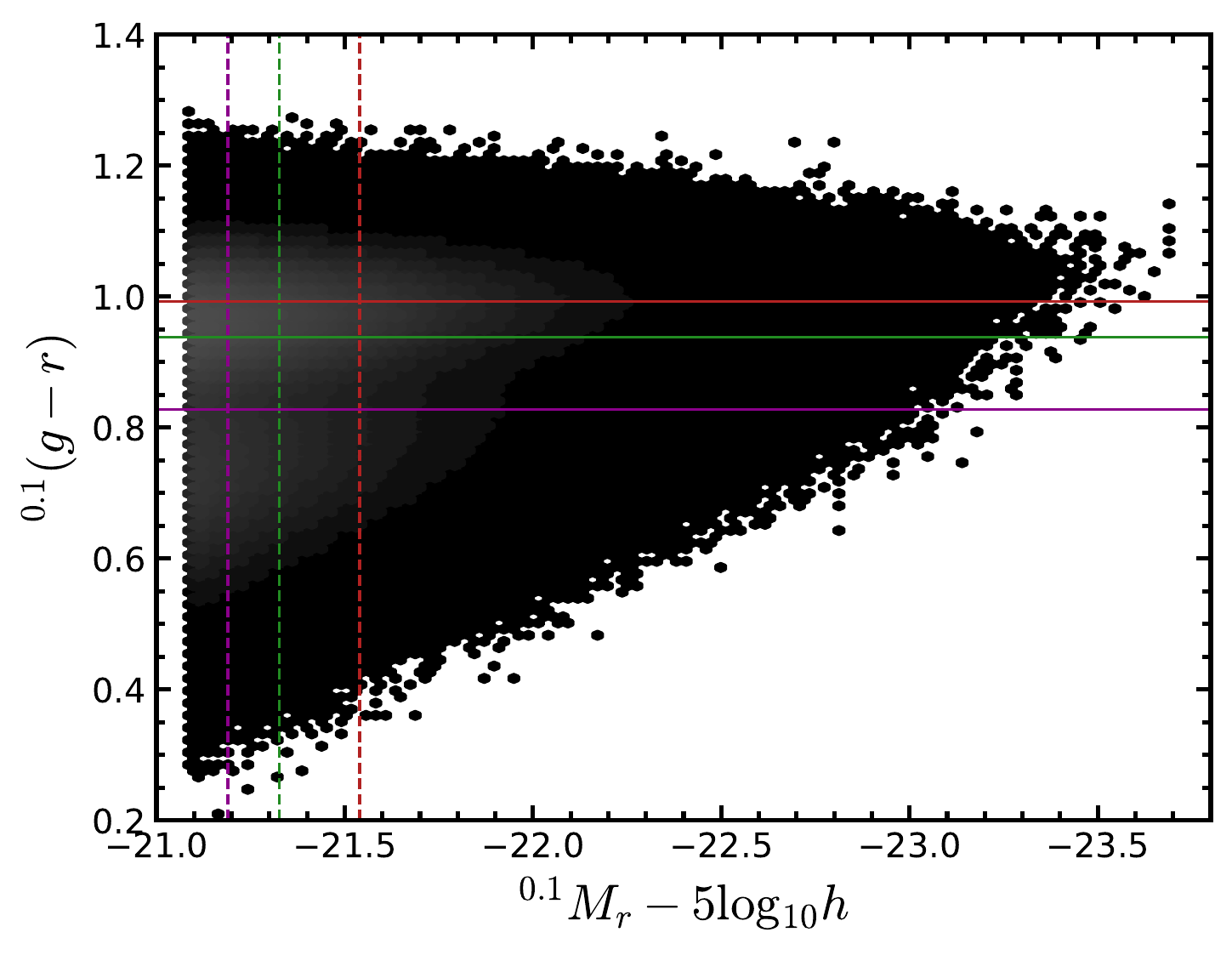}
\includegraphics[width=0.45\textwidth]{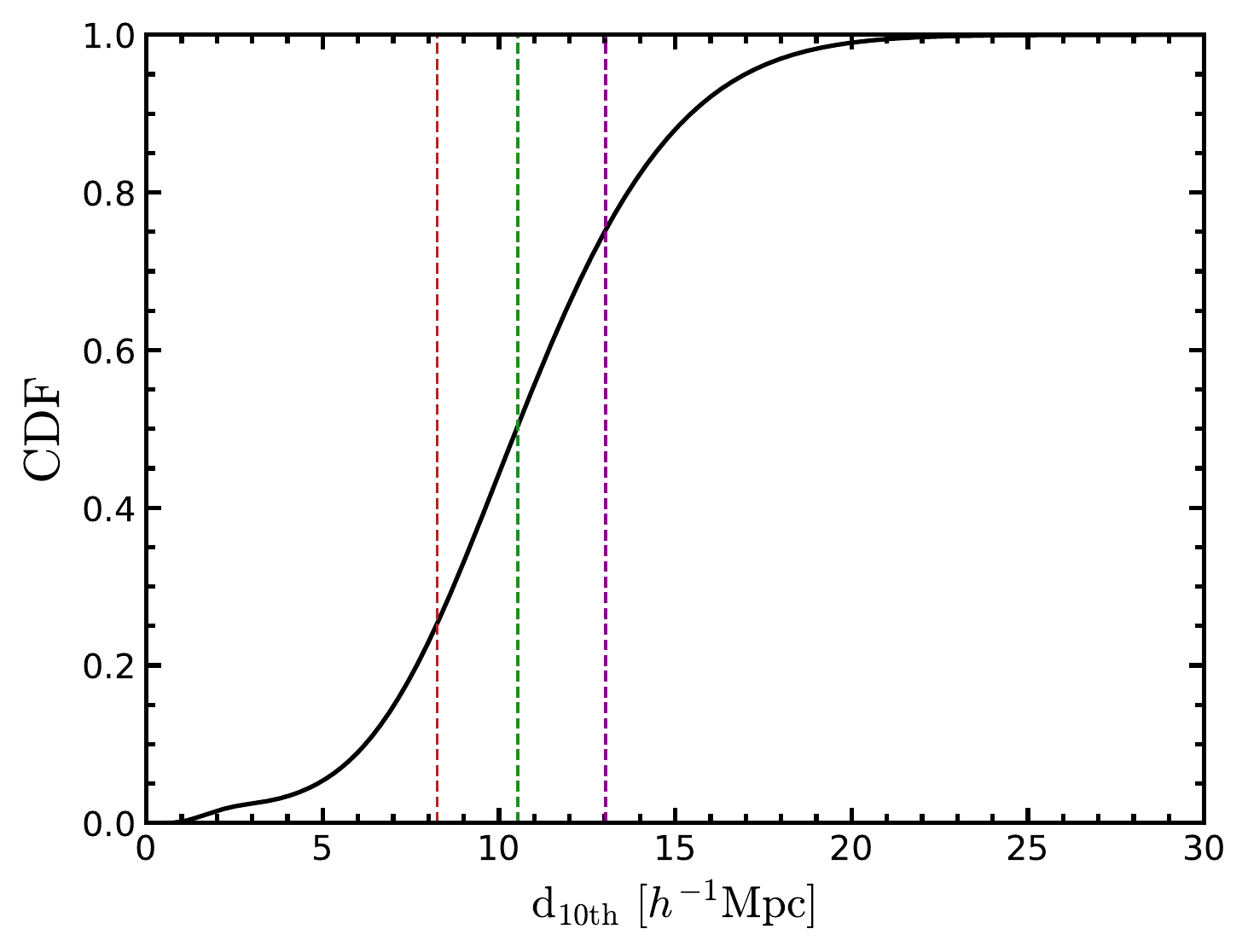}
\caption{Selection cuts applied to the full sample to get subsamples defined by magnitude, colour or density. {\it Left panel:} Colour-magnitude diagram for the full sample. Lines of different colour show the cuts in magnitude (vertical dashed lines) and colour (horizontal solid lines) applied to divide the sample into either luminosity or colour subsamples. {\it Right panel:} Cumulative distribution of the distance to the $10^{\rm th}$ nearest neighbour $(d_{\rm 10th})$; vertical dashed lines show the cuts applied to the full sample to define density quartiles.}
\label{fig:selection2}
\end{figure*}

\section{Galaxy samples and methodology}\label{sec:data}
\subsection{Galaxy catalogue}\label{sec:mock}
We build the galaxy mock catalogue using the Millennium-XXL (MXXL) dark matter only N-body  simulation output at $z=0.11$ \citep{Angulo:2012ep}. The MXXL simulation covers a comoving volume of $(3000~\Mpch)^3$ and contains $6720^3$ particles of mass $6.17 \times 10^9 \Msh$. The cosmological parameter values adopted in the MXXL simulation are the same as those used in the original Millennium simulation \citep{Springel:2005nw} and are consistent with the WMAP-1 mission results  \citep{Spergel:2003cb}: $\Omega_{\rm m}= 0.25$, $\Omega_\Lambda= 0.75$, $\sigma_8= 0.9$ , $h= 0.73$, and $n_s = 1$. The large volume of the simulation makes it ideal to study BAO.

The construction of the mock galaxy catalogue uses the halo occupation distribution (HOD) method presented by \citet[][which is based on  \citealt{Skibba:2006} and \citealt{Skibba:2009}]{Smith:2017tzz}. This method uses a set of HODs constrained using clustering measurements from the Sloan Digital Sky Survey (SDSS), for different volume limited samples, defined using $r$-band absolute magnitude cuts \citep{Zehavi2011}. These HODs are used to populate dark matter haloes in the simulation, which are identified using the \textsc{subfind} algorithm \citep{Springel2001}. We use $\Mmean$ as the halo mass definition, which corresponds to the mass enclosed by a sphere in which the average density is 200 times the mean density of the universe. Interpolating between the HODs allows each object to be assigned a magnitude, but a modification is made to the functional form of the 5-parameter HOD model to prevent the unphysical crossing of HODs for different luminosity cuts. We denote absolute magnitudes as $^{0.1}M_r-5\log_{10} h$, where the superscript 0.1 indicates that this quantity has been $k$-corrected to redshift 0.1.
The HODs are also evolved with redshift to reproduce the luminosity function measured from the SDSS at low redshifts, and the luminosity function of the GAMA survey at higher redshifts (see \citeauthor{Smith:2017tzz} for references). Each object is also assigned a $^{0.1}(g-r)$ colour, using a parametrisation of the colour-magnitude diagram. 

In \cite{Smith:2017tzz}, the HOD methodology outlined above was used to populate a halo lightcone. Here, instead of using a lightcone, we use the simulation  output at $z=0.11$. The parent galaxy catalogue has a number density of $n_g = 7.5 \times 10^{-3}\,\hMpcc$, giving 201 million galaxies in the MXXL volume, which corresponds to retaining galaxies brighter than a magnitude cut of $\Mr = -20$. 

The left panel of Fig.~\ref{fig:cLF} shows the cumulative $r$-band  luminosity function of the parent galaxy catalogue. The horizontal blue line shows a cut in number density of  $n = 1 \times 10^{-3}\hMpcc$. We will refer to this as the  ``full sample''. The HOD of the full sample is shown in the right panel of Fig.~\ref{fig:cLF}. We can see that the shape of the HOD, by construction, follows the standard functional form proposed by \citet{Zheng:2004id}, where the mean number of central galaxies per halo reaches unity above a threshold halo mass (i.e. every halo above this mass contains a central galaxy) and the occupation of satellite galaxies follows a power-law in massive haloes.

Here we study the clustering of galaxies ranked by environment (density), luminosity and colour. We divide the full sample into four equal parts, i.e., each subsample has the same number density $n_{\rm{Q}} = 2.5\times 10^{-4}\hMpcc$.

\subsubsection{Selection of samples}\label{sec:selection}
\begin{figure*}
    \centering
    \includegraphics[width=0.46\textwidth]{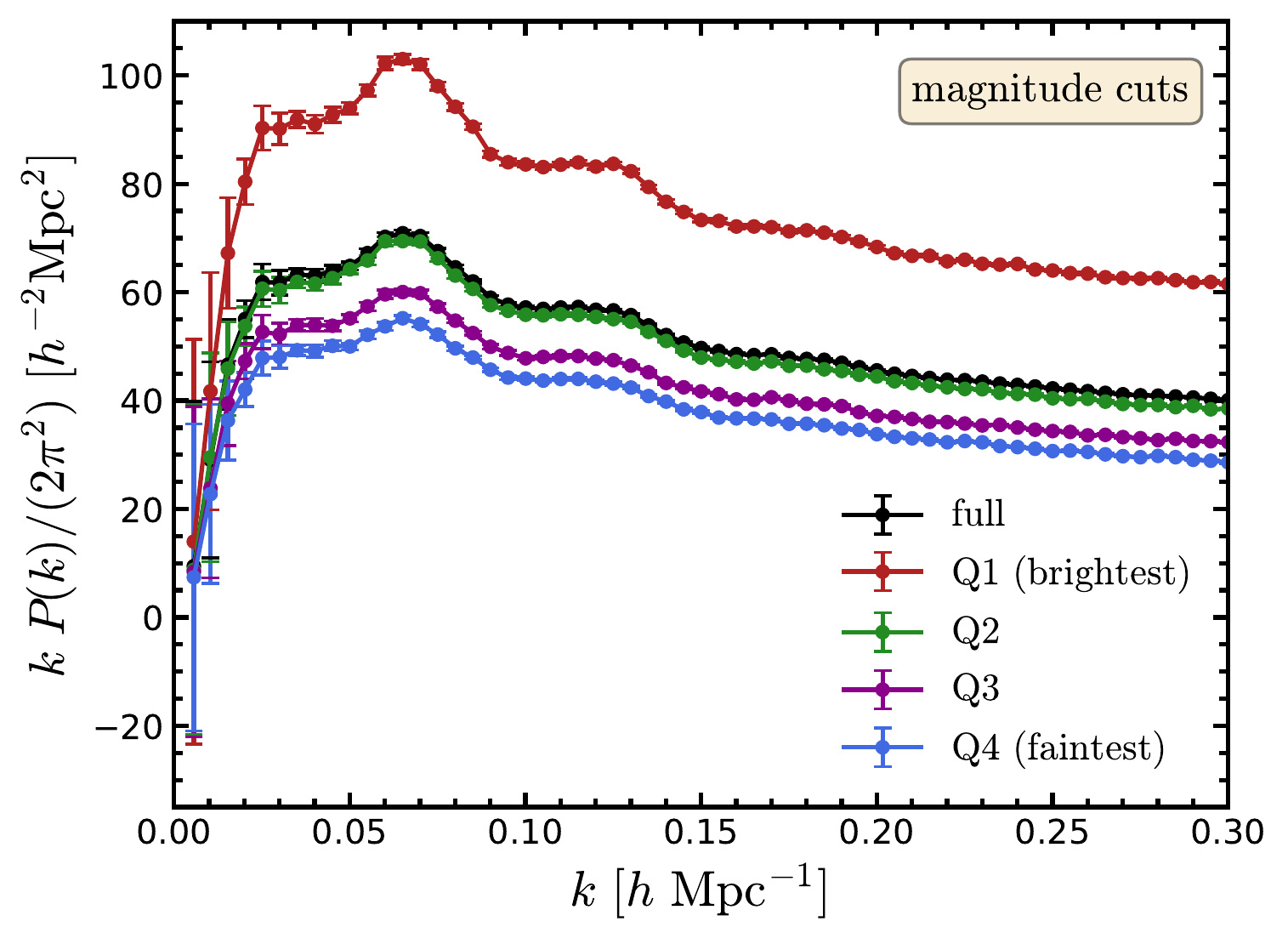}
    \includegraphics[width=0.44\textwidth]{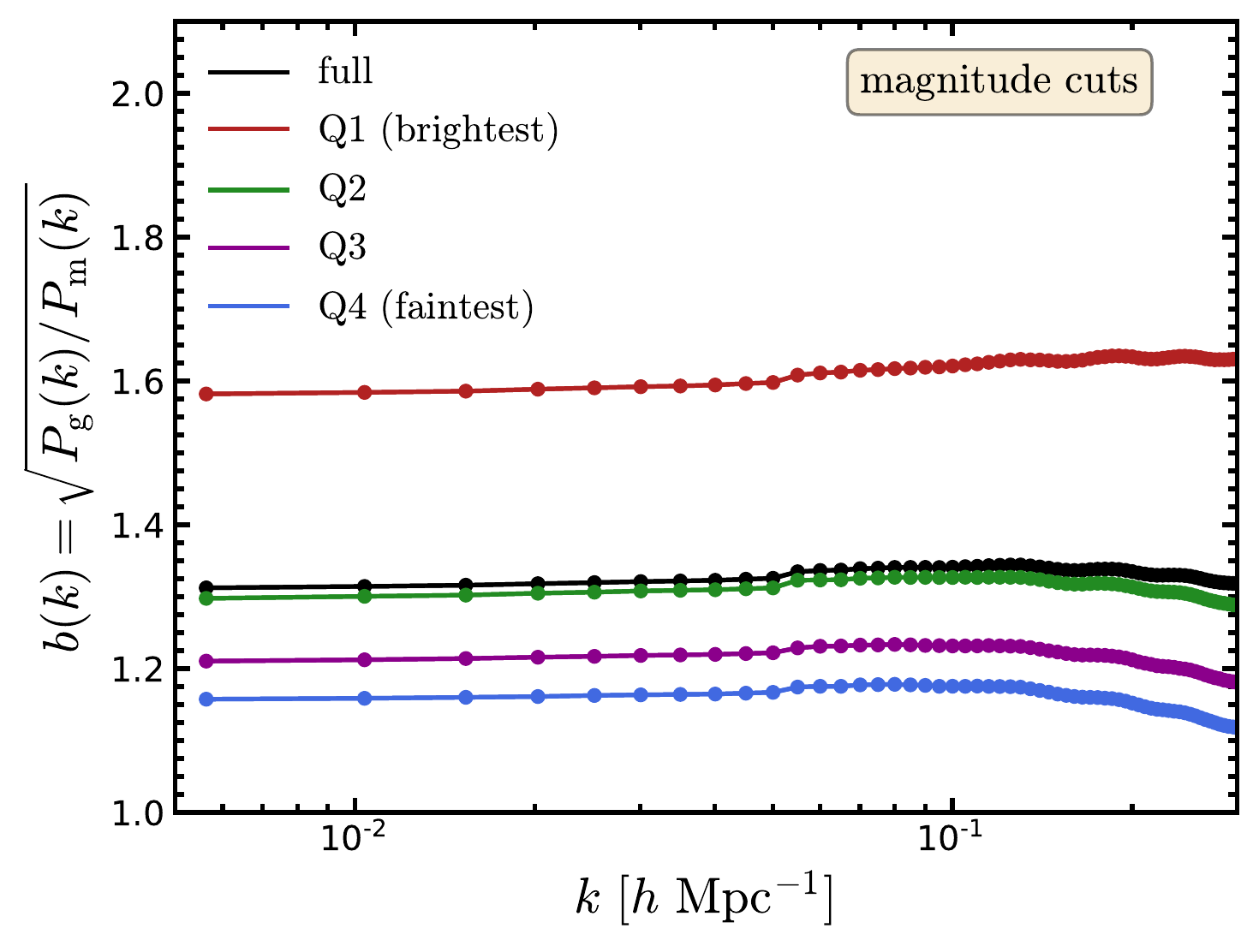}
    \includegraphics[width=0.46\textwidth]{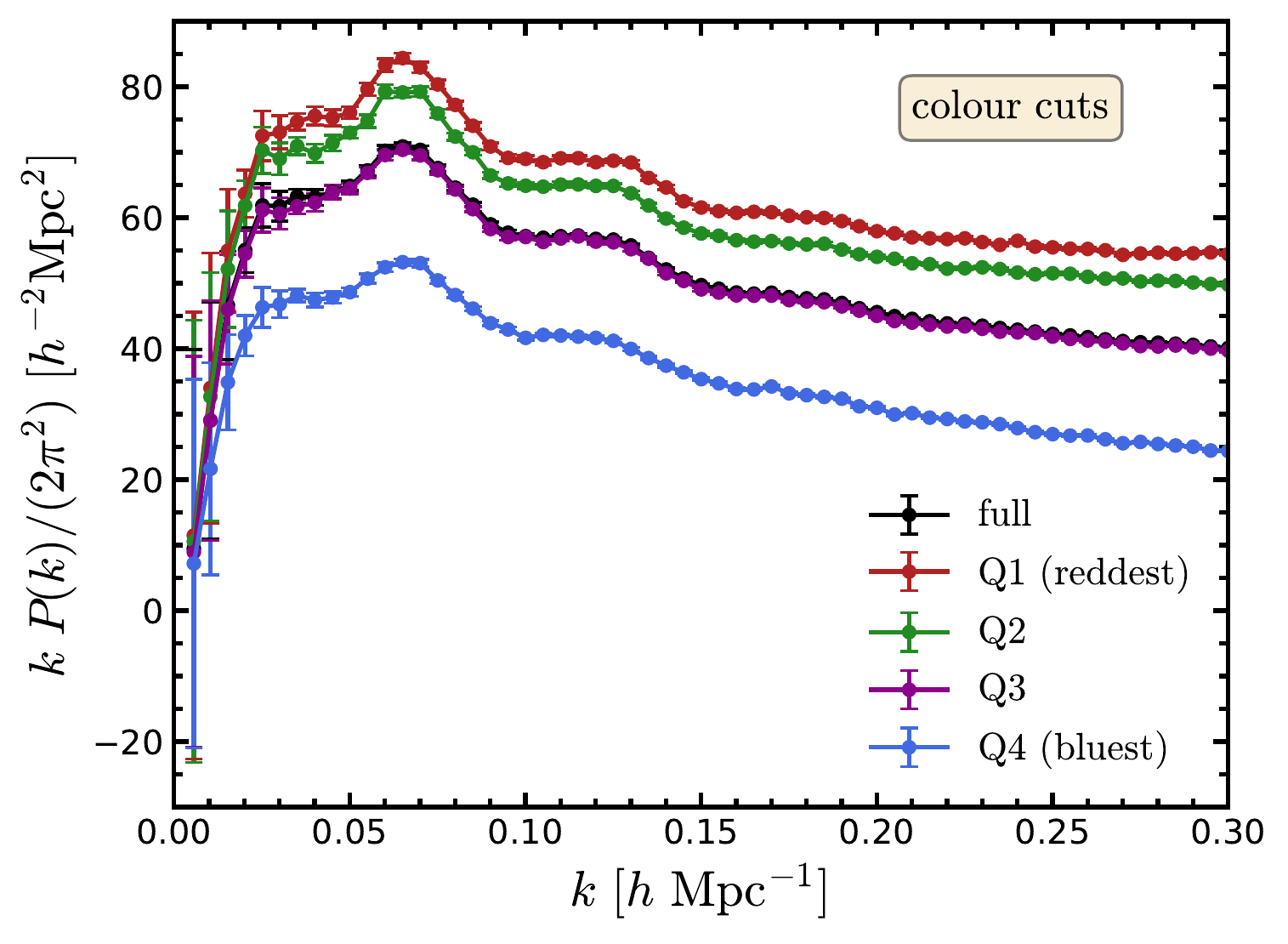}
    \includegraphics[width=0.44\textwidth]{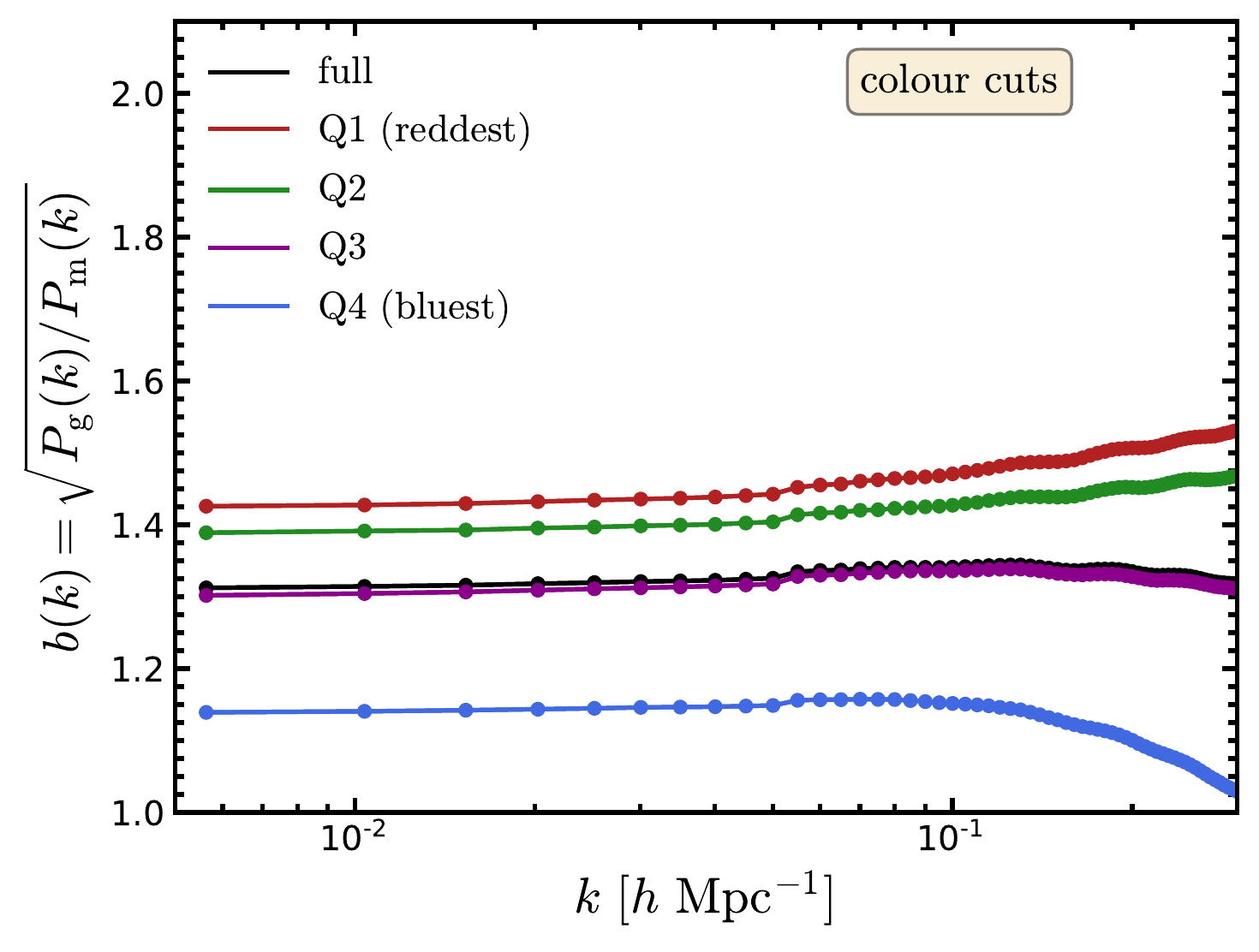}
    \includegraphics[width=0.465\textwidth]{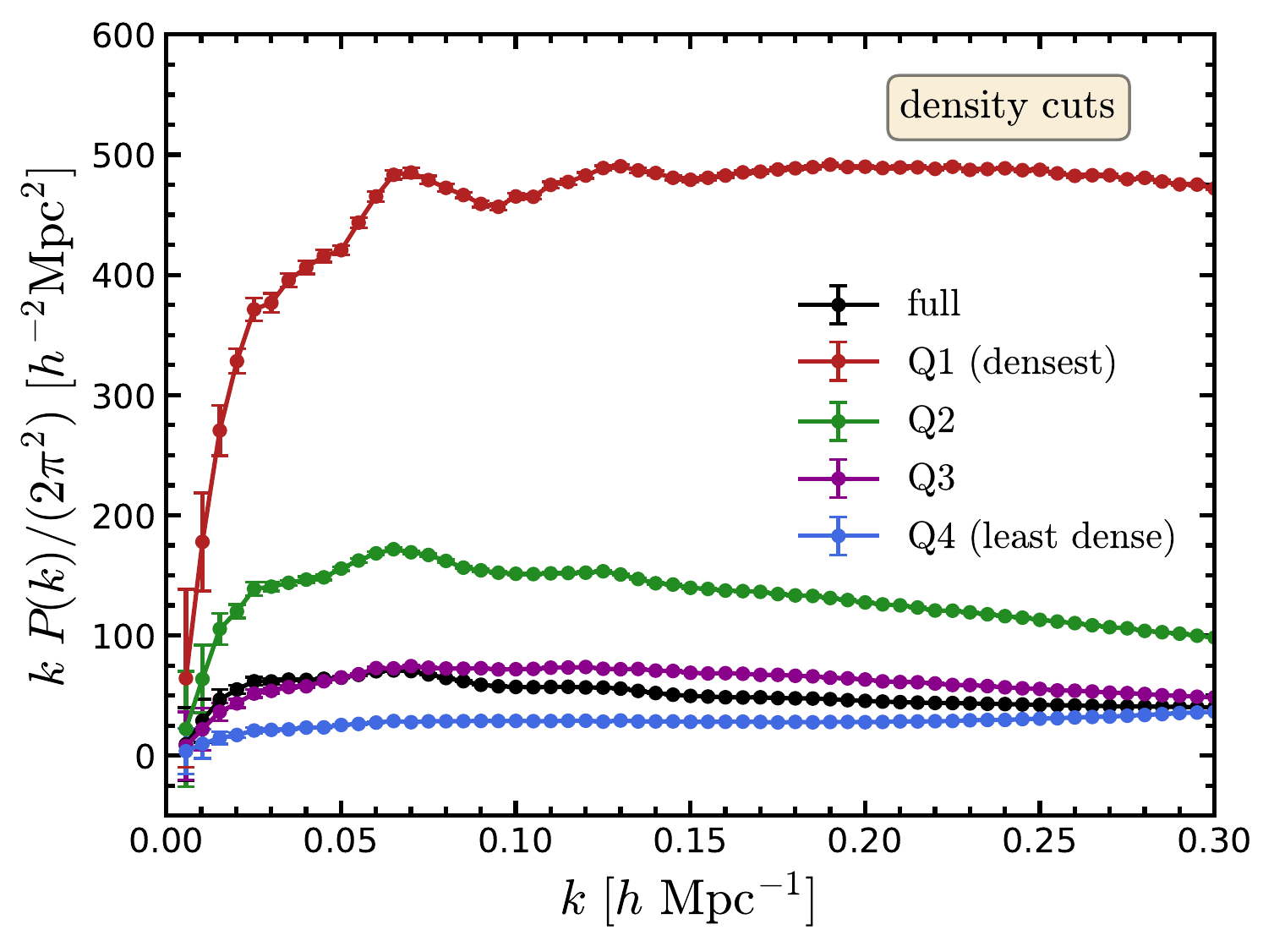}
    \includegraphics[width=0.44\textwidth]{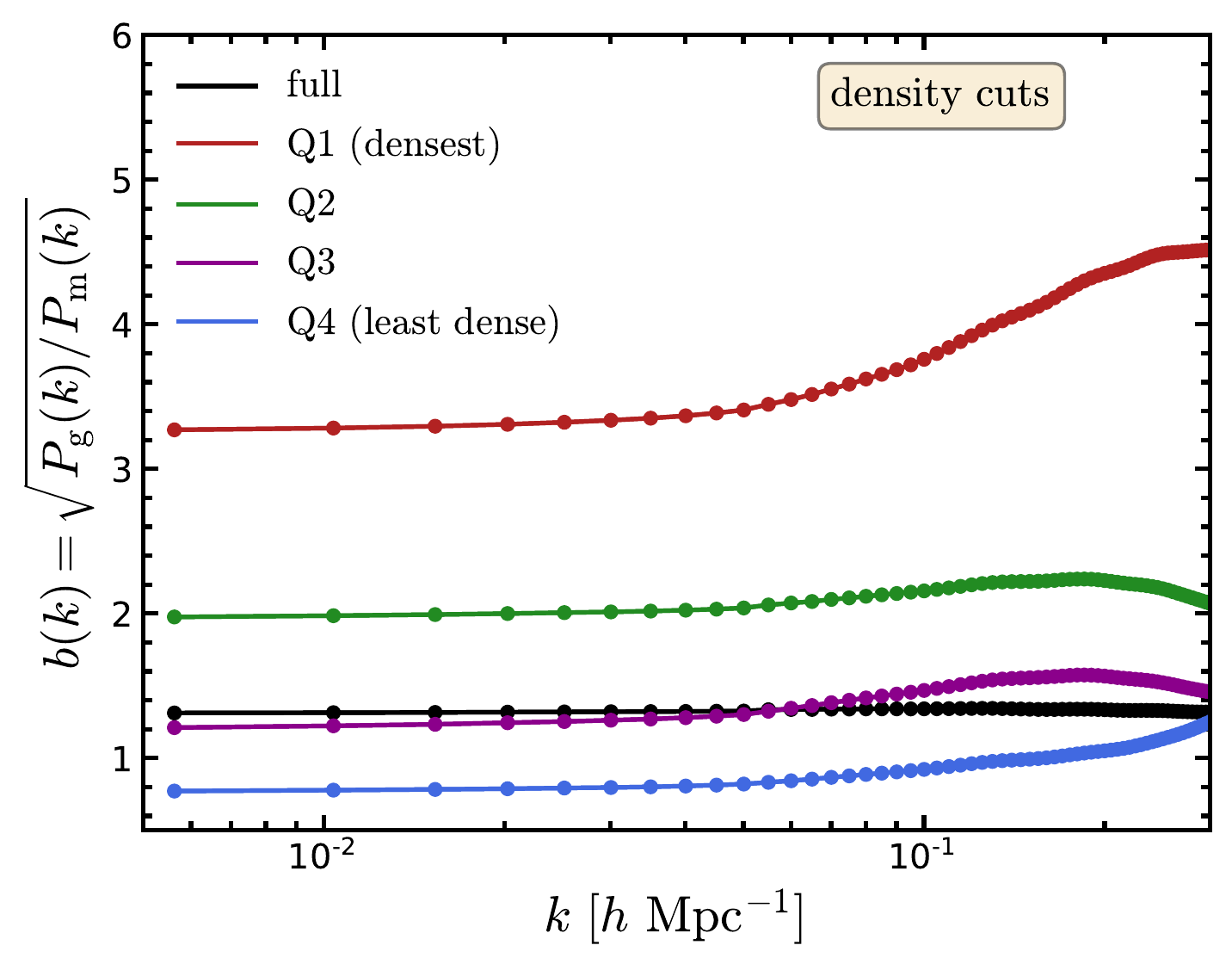}
    \caption{The measured power spectrum, $P(k)$ ({\it left column}), and the galaxy bias, $b(k) = \sqrt{P_{\rm g}(k)/P_{\rm m}(k)}$ ({\it right column}), of the four subsamples for each galaxy selection: magnitude ({\it upper panels}), colour ({\it middle panels}) and density ({\it lower panels}). Different colours represent different subsamples as labelled: red (Q1), green (Q2), magenta (Q3) and blue (Q4). In each panel we show the measured power spectrum ({\it left}) and galaxy bias ({\it right}) from the full sample (black solid points) for comparison. Note that the $y$-axis range plotted is different in each panel.}
    \label{fig:Pk_samples}
\end{figure*}

\begin{figure}
    \centering
    \includegraphics[width=0.45\textwidth]{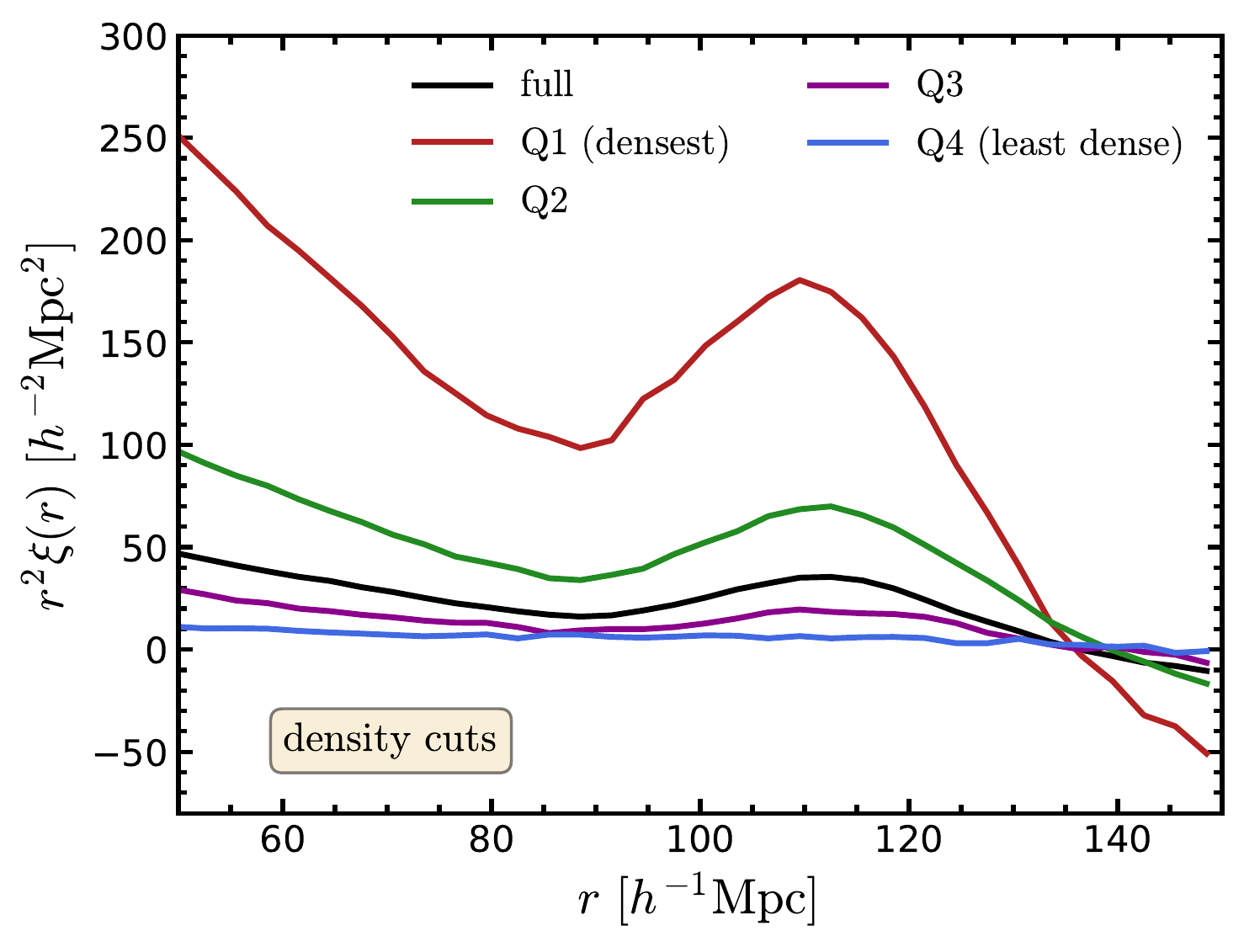}
    \caption{The correlation function of the galaxy samples defined by density, plotted as $r^2 \xi(r)$, on a linear-linear scale. The black line shows the correlation function measured for the full galaxy sample and the coloured lines show the clustering for the subsample quartiles ranked by density, as labelled.}
	\label{fig:xi_dens_all}
\end{figure}

We first select galaxies by luminosity, retaining those which satisfy cuts in magnitude. The vertical lines in the left panel of Fig.~\ref{fig:selection2} show the magnitude bins used to define the luminosity quartiles: the $\rm{Q}_1$ subsample corresponds to the brightest 25 per cent of galaxies while $\rm{Q}_4$ is the subsample with the 25 per cent faintest galaxies. 
We next apply the colour cuts listed in Table~\ref{tab:cuts} to define the colour subsamples, shown by the horizontal lines in the left panel of Fig.~\ref{fig:selection2}, where $\rm{Q}_1$ is the subsample with the 25 per cent reddest galaxies and $\rm{Q}_4$ contains the 25 per cent bluest galaxies.
Finally, to define samples by environment we apply a cut in local density. The local density is estimated using the distance to the 10$^{\rm th}$ nearest neighbour, $d_\mathrm{10th}$, and the galaxies are ranked by this property. 
The right panel of Fig.~\ref{fig:selection2} shows the cumulative distribution function of $d_\mathrm{10th}$ where the vertical dotted lines show the quartiles. The first quartile of the sample $(\rm{Q}_1)$ contains the 25 per cent of  galaxies in the densest environments (i.e. those  with the smallest values of $d_\mathrm{10th}$) and $\rm{Q}_4$ is the subsample with the 25 per cent of the galaxies in the least-dense environments. $\rm{Q}_2$ and $\rm{Q}_3$ are the subsamples in intermediate density regions. The values of $d_\mathrm{10th}$ used to define the density samples are listed in Table~\ref{tab:cuts}.

\begin{table}
    \centering
    \caption{The selection cuts applied to define galaxy subsamples in terms of luminosity ($^{0.1}M_r - 5\mathrm{log}_{10} h$), colour ($^{0.1}(g-r)$) or density ($d_\mathrm{10th}/ \Mpch$).}
\begin{tabular}{lccccc}
\multicolumn{6}{c}{}\\ \hline
\multicolumn{6}{c}{$^{0.1}M_r - 5\mathrm{log}_{10} h$}                                                     \\ \hline
\multicolumn{1}{c}{} & $\mathrm{full}$ & $\mathrm{Q_1}$ & $\mathrm{Q_2}$ & $\mathrm{Q_3}$ & $\mathrm{Q_4}$ \\
bright limit                  & -23.70          & -23.70         & -21.52         & -21.32         & -21.18         \\
faint limit                  & -21.08          & -21.53         & -21.33         & -21.19         & -21.08         \\ \hline
\multicolumn{6}{c}{$^{0.1}(g-r)$}                                                                          \\ \hline
\multicolumn{1}{c}{} & $\mathrm{full}$ & $\mathrm{Q_1}$ & $\mathrm{Q_2}$ & $\mathrm{Q_3}$ & $\mathrm{Q_4}$ \\
blue limit                  & 0.21            & 1.00           & 0.94           & 0.83           & 0.21           \\
red limit                  & 1.28            & 1.28           & 0.99           & 0.93           & 0.82           \\ \hline
\multicolumn{6}{c}{$d_\mathrm{10th}/ \Mpch$}                                                               \\ \hline
                     & $\mathrm{full}$ & $\mathrm{Q_1}$ & $\mathrm{Q_2}$ & $\mathrm{Q_3}$ & $\mathrm{Q_4}$ \\
most dense                   & 0.26            & 0.26           & 8.26           & 10.54          & 13.04          \\
least dense                   & 33.95           & 8.25           & 10.53          & 13.03          & 33.95         
\end{tabular}
\label{tab:cuts}
\end{table}

\begin{figure*}
    \centering
    \includegraphics[width=0.44\textwidth]{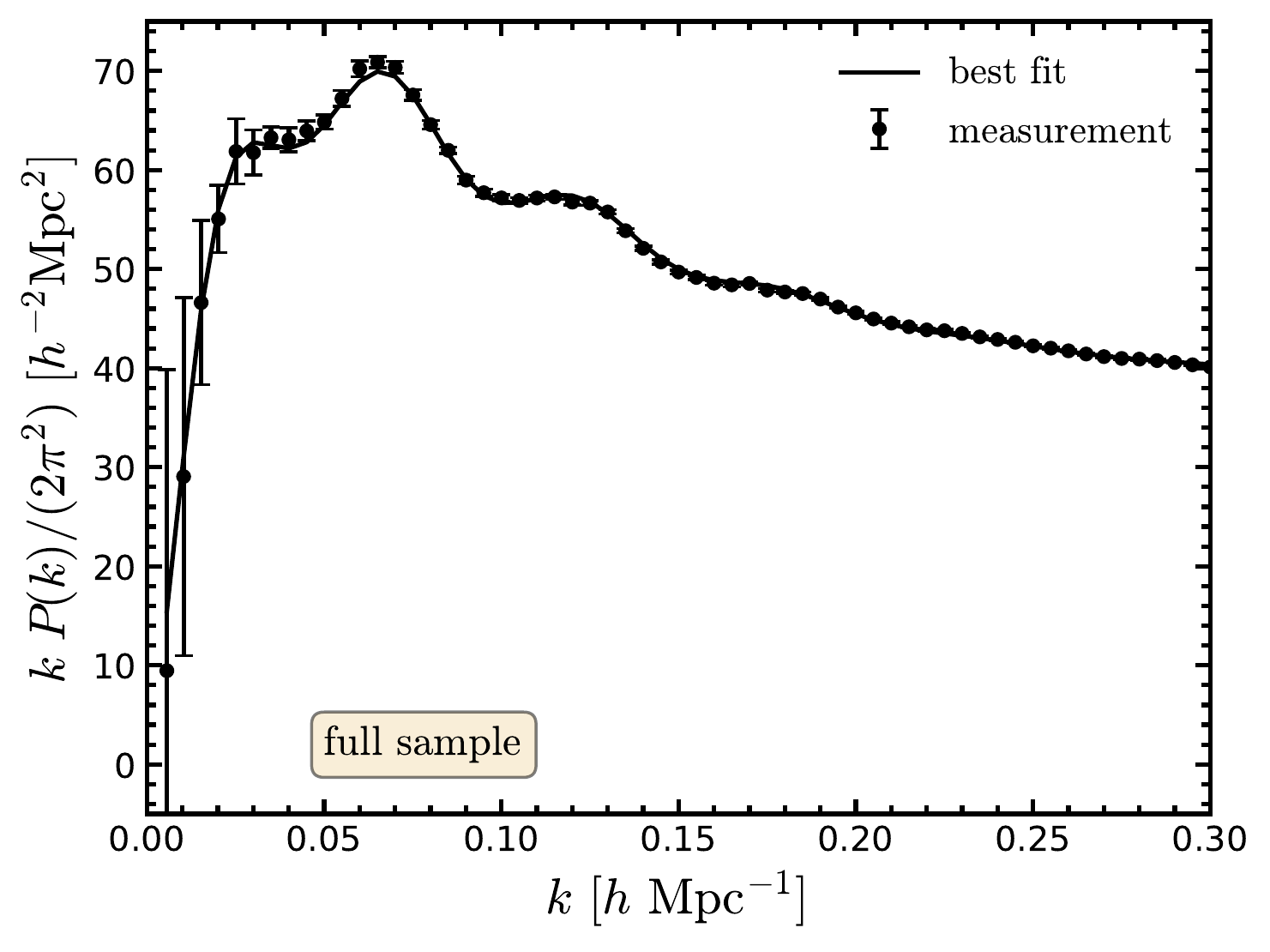}
    \includegraphics[width=0.45\textwidth]{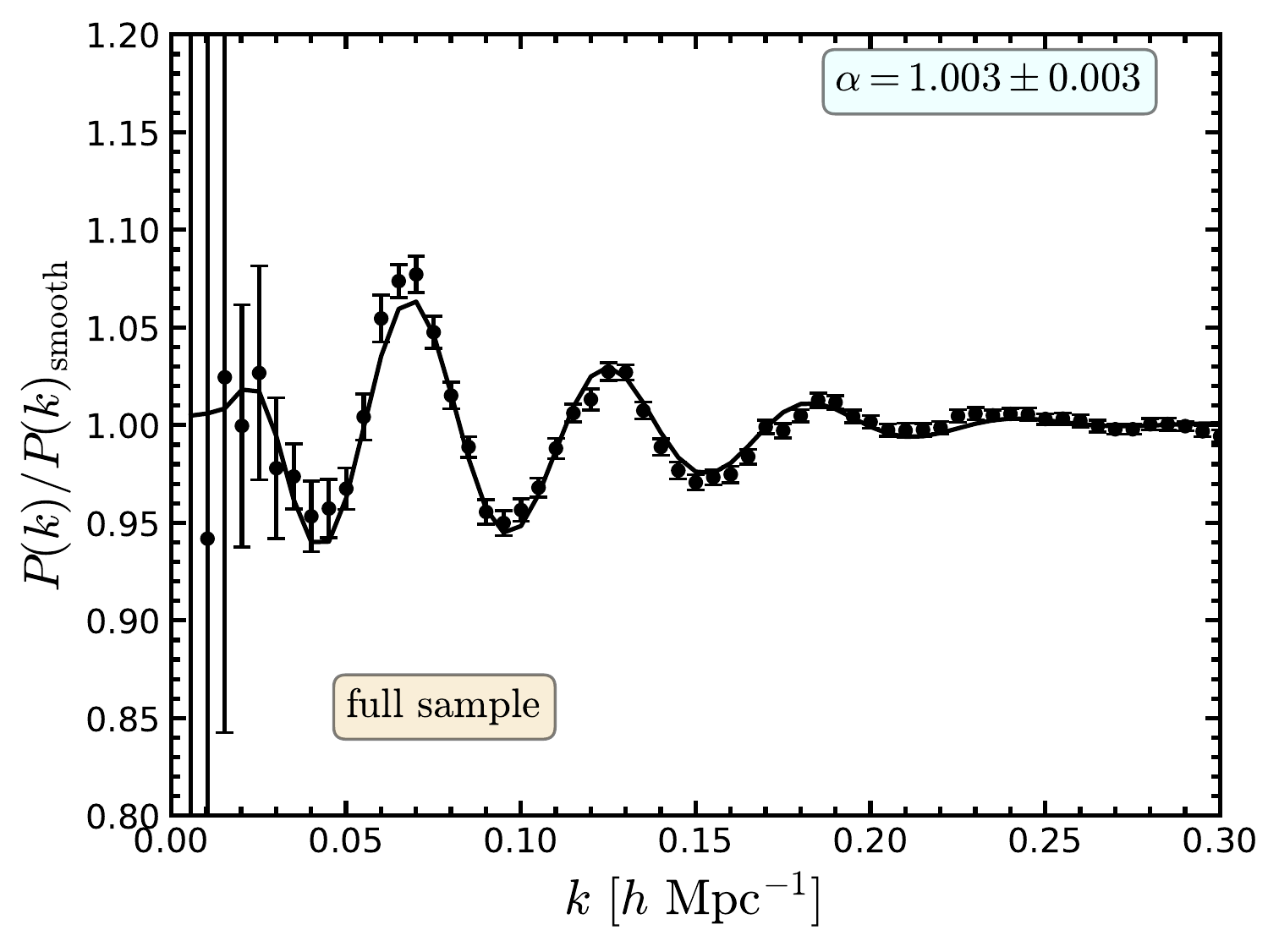}
    \caption{{\it Left panel:} The measured power spectrum, $P(k)$, (points with error-bars) and the best-fitting model (solid curve) for the full galaxy sample. {\it Right panel:} The same as the left panel but now the power spectrum is plotted divided by the smooth (no-wiggle) component of the best-fitting model. This panel highlights the BAO signature, which corresponds to the oscillations of the curve. The upper label in the right panel indicates how accurately we can measure the BAO scale, as parametrised in terms of the $\alpha$ dilation parameters (see main text for details).}
    \label{fig:Pk_n1}
\end{figure*}

The left panels of Fig.~\ref{fig:Pk_samples} display the real-space power spectrum measured from galaxy samples ranked by magnitude (top), colour (middle) and density (bottom) as listed in Table~\ref{tab:cuts}. The black points in each panel correspond to the power spectrum of the full galaxy sample, in which we can clearly see the BAO wiggles in Fourier space. 

It is evident when comparing measurements across different selections that the $\rm{Q}_1$ subsamples (i.e. the brightest, reddest and densest galaxies) shown in Fig.~\ref{fig:Pk_samples} are more clustered and therefore show a higher galaxy bias than the overall sample. It is interesting to see that the magnitude subsample ${\rm Q}_2$ (green solid line in the top-left panel) and the colour subsample ${\rm Q}_3$ (magenta solid line in the middle-left panel) have almost the same clustering amplitude as the full sample. The BAO peaks measured from the densest subsample are {\it significantly} stronger than those seen in the measurements made from the other samples (note the y-axis range plotted is different in each panel). The BAO peaks are barely visible for the least dense sample ($\rm{Q}_4$, bottom-left panel).

The right column of Fig.~\ref{fig:Pk_samples} shows the galaxy bias for every subsample. The bias is obtained as
\begin{equation}
    b(k) = \sqrt{\frac{P_{\rm g}(k)}{P_{\rm m}(k)}}\,,
\end{equation}
where $P_{\rm g}(k)$ is the measured galaxy power spectrum for each subsample (the same as shown in the left panels of Fig.~\ref{fig:Pk_samples}) and $P_{\rm m}(k)$ is the non-linear dark matter power spectrum at $z=0.11$. We can see that the galaxy bias inferred for each subsample is constant on large scales ($k \lesssim 0.1 \hMpc$). The scale dependence becomes evident at higher wavenumbers, with the bias increasing (e.g. for the reddest, densest and brightest subsamples) and decreasing for  the bluest and faintest subsamples. The scale dependence of the bias is particularly strong for the subsamples defined by local density. 

In Fig.~\ref{fig:xi_dens_all} we show the two-point correlation function on  scales around $r\sim100\Mpch$ that correspond to the location of the BAO peak. The figure shows the two-point correlation for the full sample of galaxies (black line) as well as for the density-selected quartiles. Similar trends are observed for the magnitude- and colour-selected subsamples, which, for brevity, we do not show.
As expected from our power spectrum results, the ${\rm Q}_1$ density-subsample displays the strongest clustering, i.e. galaxies in the densest regions are more likely to reside in more massive haloes, which are more biased, and hence we measure a higher clustering amplitude for this subsample. The BAO wiggles are clearer for this sample in the power spectrum and the BAO peak is stronger in the correlation function (see bottom-left panel of Fig.~\ref{fig:Pk_samples} and Fig.~\ref{fig:xi_dens_all}). We also note that non-linear effects are more evident in the densest sample on small-scales. There is an increase in the power for scales $k > 0.15 \hMpc$, and a steeper slope in the correlation function at $r < 70 \Mpch$ (Fig.~\ref{fig:xi_dens_all}). We note that the BAO feature is slightly shifted to smaller  scales in the highest density  subsample, i.e. the position of the peak is moved to higher $k$ values in the power spectrum and to lower $r$ values in the correlation function \citep[as predicted by][]{Neyrinck:2016pfm}.

\subsection{BAO model}\label{sec:model}
Here, we measure the BAO scale in the power spectrum of galaxies. To do this, we follow a similar approach to that presented by \citet[][see also \citealt{Eisenstein2007}]{Ross:2014qpa}. We start by modelling the power spectrum as the product of a smooth component and the BAO signal. That is, we write the model power spectrum, $P_{\rm fit}(k)$, as
\begin{equation}\label{eq:Pfit}
    P_{\rm fit}(k) = P_{\rm sm}(k) {\rm O}_{\rm damp}(k/\alpha)\,,    
\end{equation}
where $P_{\rm sm}(k)$ is a smooth power spectrum, i.e., without any BAO feature, and ${\rm O}_{\rm damp}(k/\alpha)$ represents the damped BAO signal. The damping factor is parametrised in terms of the $\alpha$ dilation parameter that characterises any shift in the position of the BAO peak in the measured power spectrum compared to the model; if $\alpha > 1$ the peak is moved to smaller scales, while $\alpha < 1$ moves the peak to larger scales \citep{Angulo:2008,Anderson:2013zyy,Ross:2014qpa}. This template can be used to analyse the galaxy power spectrum in both real and redshift space.

We model the smooth power spectrum component as
\begin{equation}\label{eq:Psm}
    P_{\rm sm}(k) = B^2_p P_{\rm nw}(k) + A_1k + A_2 + \frac{A_3}{k}\,,
\end{equation}
where $P_{\rm nw}(k)$ is a smooth ``no-wiggle'' template obtained using the fitting formula of \citet{Eisenstein:1997ik}, $B_p$ is a large-scale bias parameter, and $A_1$, $A_2$ and $A_3$ are further free parameters. This functional form is similar to that used by \citet{Ross:2014qpa}, however with fewer (4 instead of 6) free parameters. We find that this function provides a very good description of the non-linear galaxy power spectrum down to $k=0.3\hMpc$.

The oscillatory component of the power spectrum is given by,
\begin{equation}\label{eq:Odamp}
    {\rm O}_{\rm damp}(k) = 1 + \left({\rm O}_{\rm lin}(k) - 1\right)e^{-\frac{1}{2}k^2\Sigma^2_{\rm nl} }\,,
\end{equation}
where $\Sigma_{\rm nl}$ is a damping parameter and ${\rm O}_{\rm lin}(k)$ is the ratio between the linear power spectrum and the smooth no-wiggle power spectrum, i.e. $P_{\rm lin}(k)/P_{\rm nw}(k)$.

We estimate the analytical power spectrum with the {\sc Nbodykit} toolkit \citep{Hand:2017pqn}, using the {\sc class} transfer function for the linear power spectrum \citep{Blas:2011rf,Lesgourgues:2011re} and the analytical approximation of \citet{Eisenstein:1997ik} for the no-wiggle power spectrum in Eqs.~\eqref{eq:Psm} and \eqref{eq:Odamp}. We also use {\sc Nbodykit} to measure the power spectrum from the simulation outputs for wavenumbers between $0.0025 < k/[\hMpc] < 0.3$ using bins with width $\Delta k = 0.005 \hMpc$. 

To measure the position of the BAO peak, we fit the measured real-space power spectrum of our subsamples to the model given by Eq.~\eqref{eq:Pfit} and extract information about the dilation parameter $\alpha$.
To obtain the best-fitting $\alpha$ value, we use Bayesian statistics and maximise the likelihood, $\mathcal{L}\propto \exp(-\chi^2/2)$ by fitting the measurements from the galaxy samples on scales with $k < 0.3 \hMpc$. We estimate errors on the measurements using 8 jackknife partitions along each coordinate of the simulation box \citep{Norberg:2009}. To find the best-fitting $\alpha$ value and its confidence levels we use the Monte Carlo Markov Chain technique implemented in the {\sc emcee} python package \citep{emcee:2013}.

For the density-selected samples, the measured power spectrum cannot be adequately described by Eq.~\eqref{eq:Psm}. We reduce the scale dependence of the power spectrum by defining a $k$-space window flattening function, $B_{k-{\rm window}}(k)$, which is the ratio between the power spectrum measured for one of the density quartile samples, divided by the power spectrum of the full sample. A similar approach was employed in \cite{Angulo:2008}. In this exercise, the two power spectra in question are first rebinned into broader $k$-bins ($\Delta k=0.1\hMpc$) before taking the ratio. The measured power spectrum is then divided by the flattening function, $B_{k-{\rm window}}(k)$, before being fitted. The window width is chosen to be larger than the scale of the BAO oscillations, and thus should be largely insensitive to the presence of the BAO signal. We have tested that this procedure does not introduce biases in $\alpha$ or in its uncertainties by testing that the luminosity- and colour-selected quartiles return the same $\alpha$ best fit values when fitting directly the sub-sample power spectrum or the one normalised using the flattening function we just discussed.

\section{Galaxy clustering}\label{sec:clustering}
\subsection{Measuring BAO positions}\label{sec:fits}
\begin{figure*}
    \centering
    \includegraphics[width=0.33\textwidth]{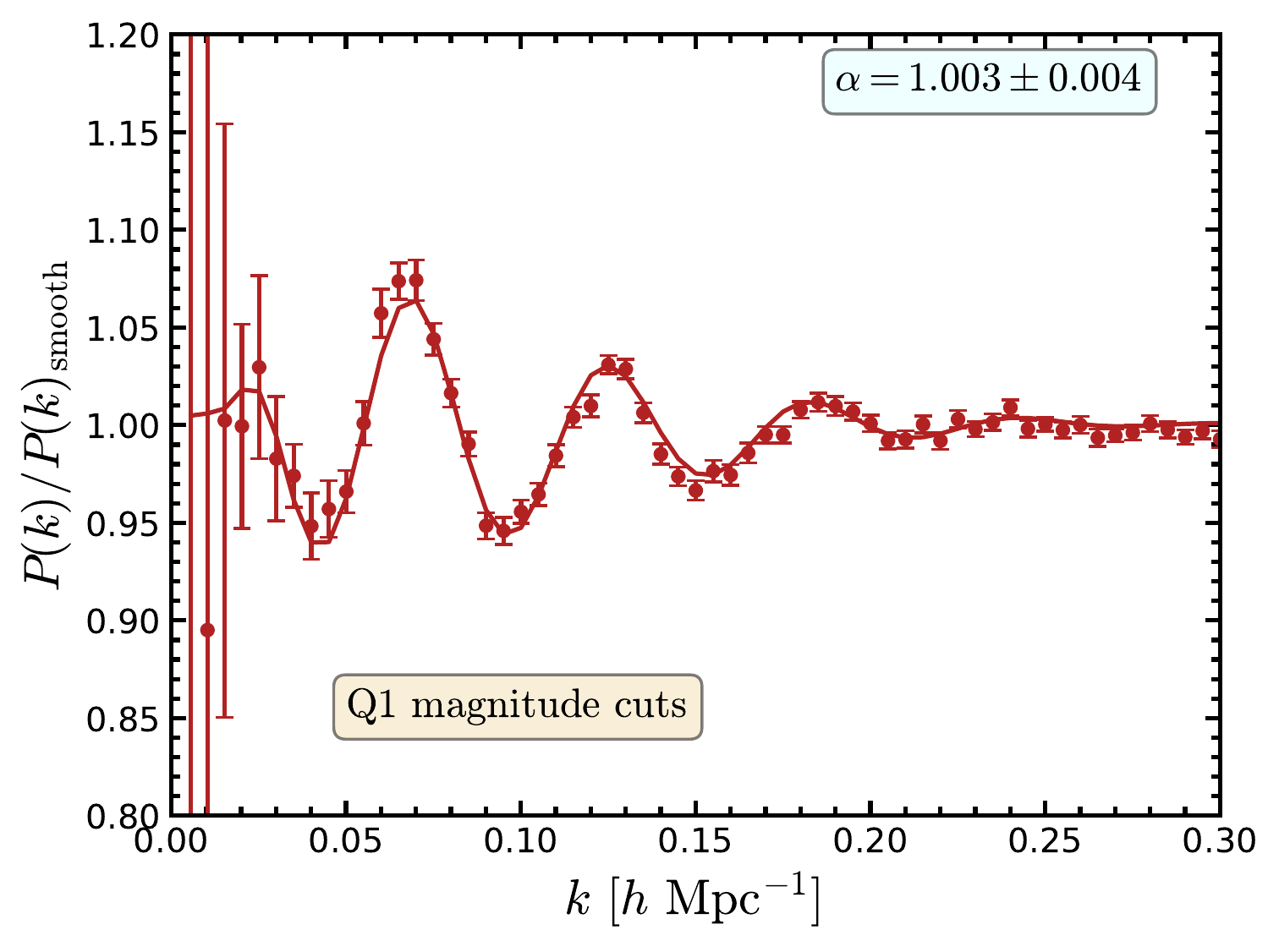}
    \includegraphics[width=0.33\textwidth]{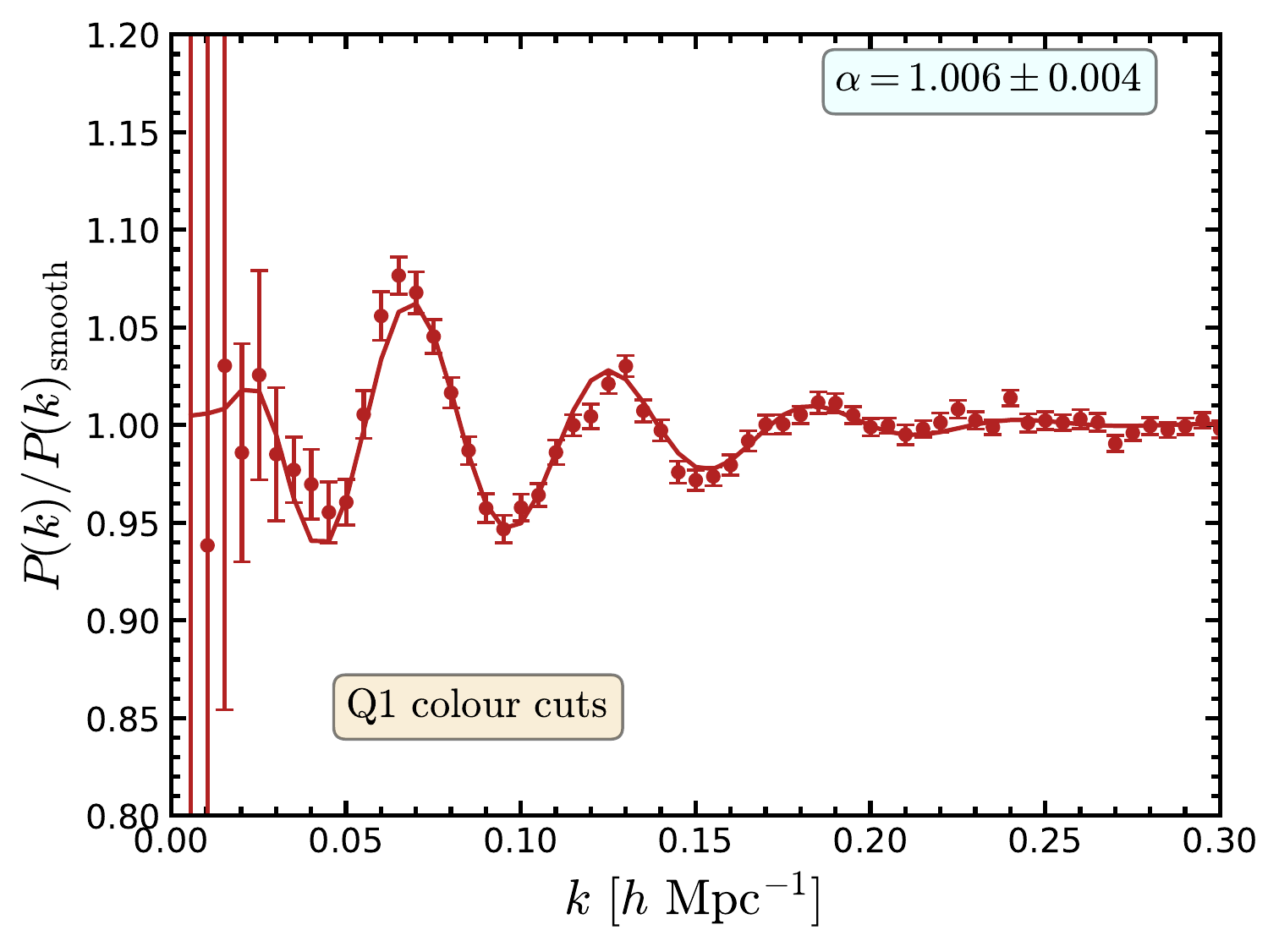}
    \includegraphics[width=0.33\textwidth]{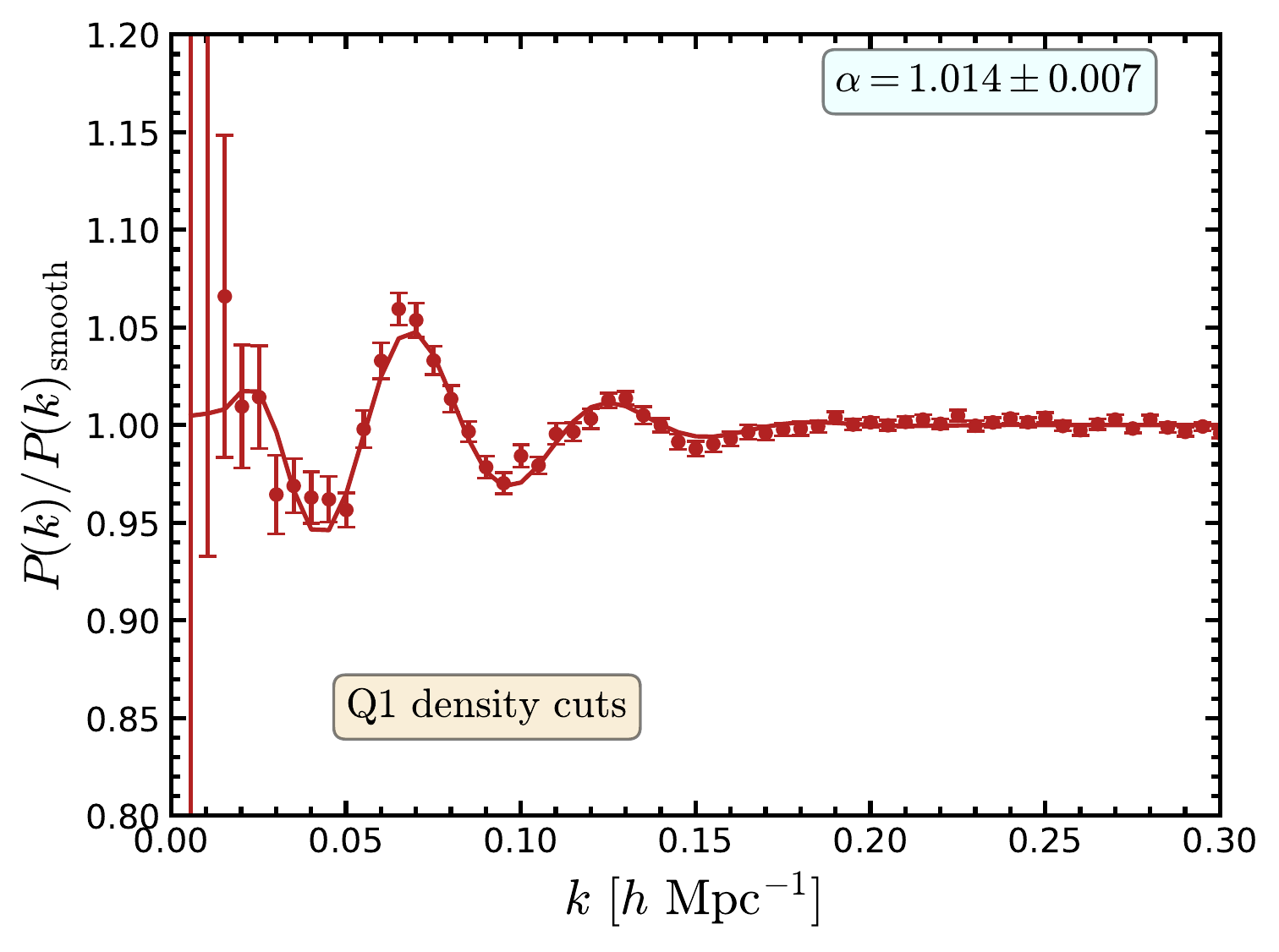}
    \includegraphics[width=0.33\textwidth]{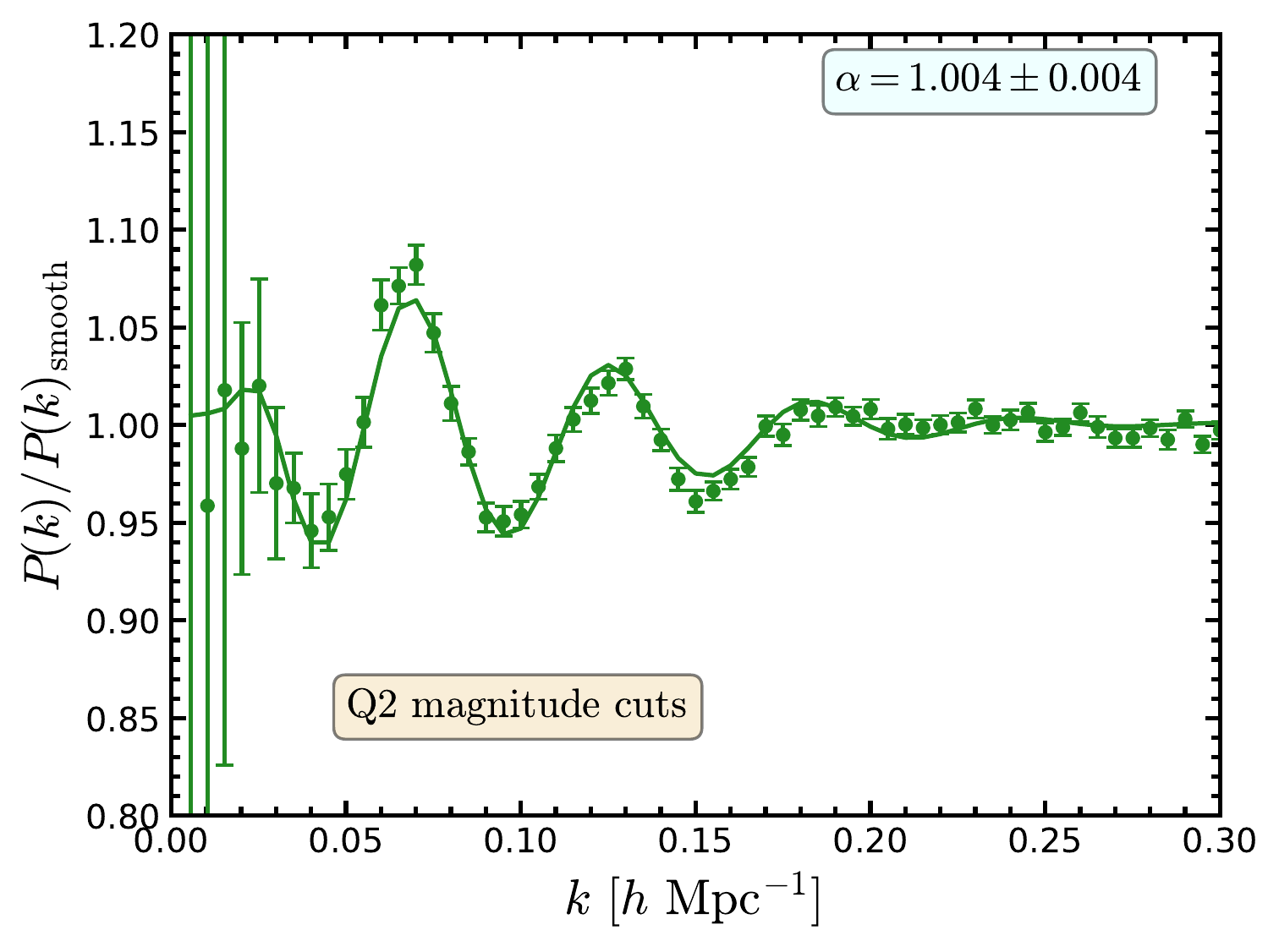}
    \includegraphics[width=0.33\textwidth]{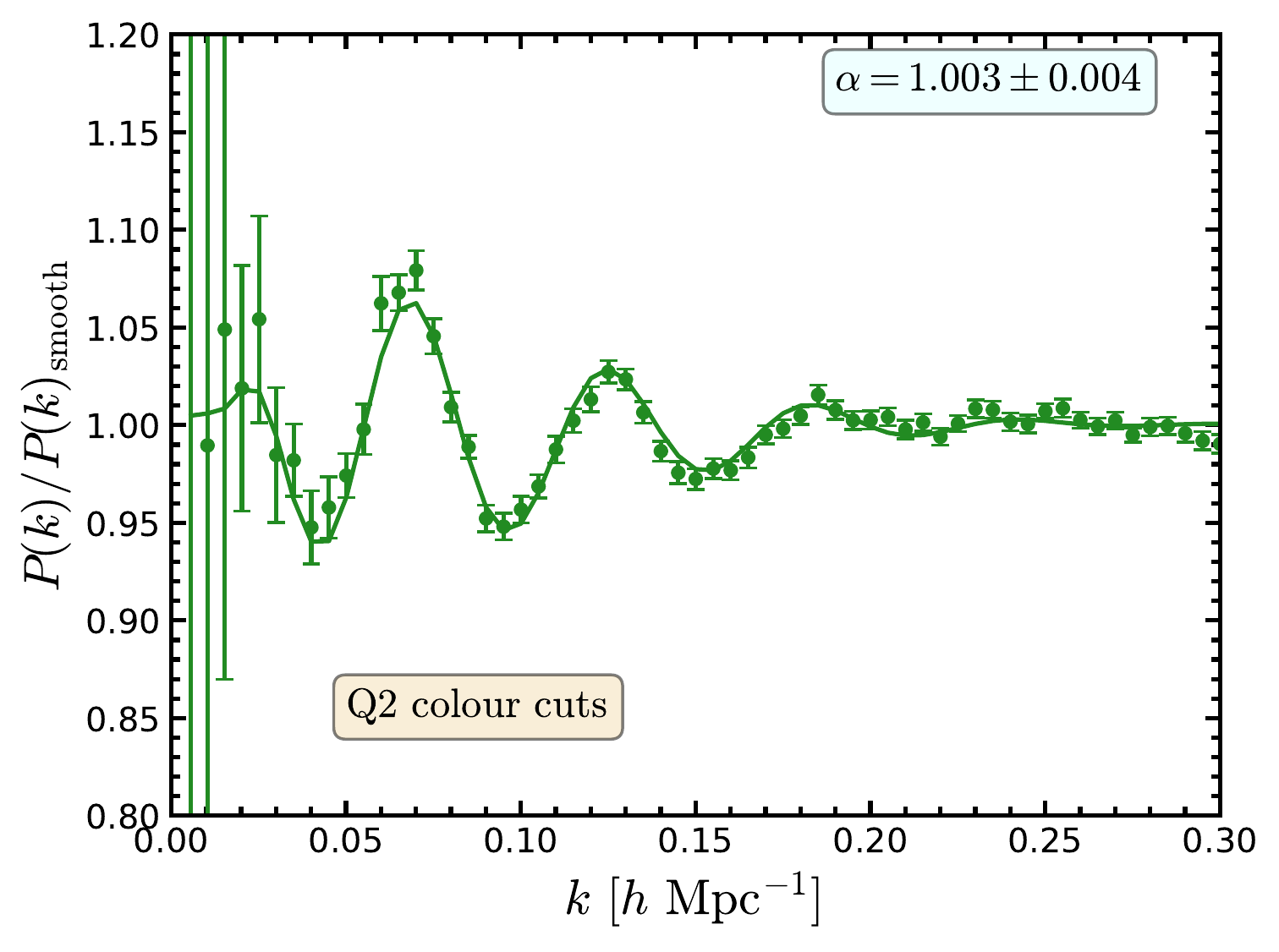}
    \includegraphics[width=0.33\textwidth]{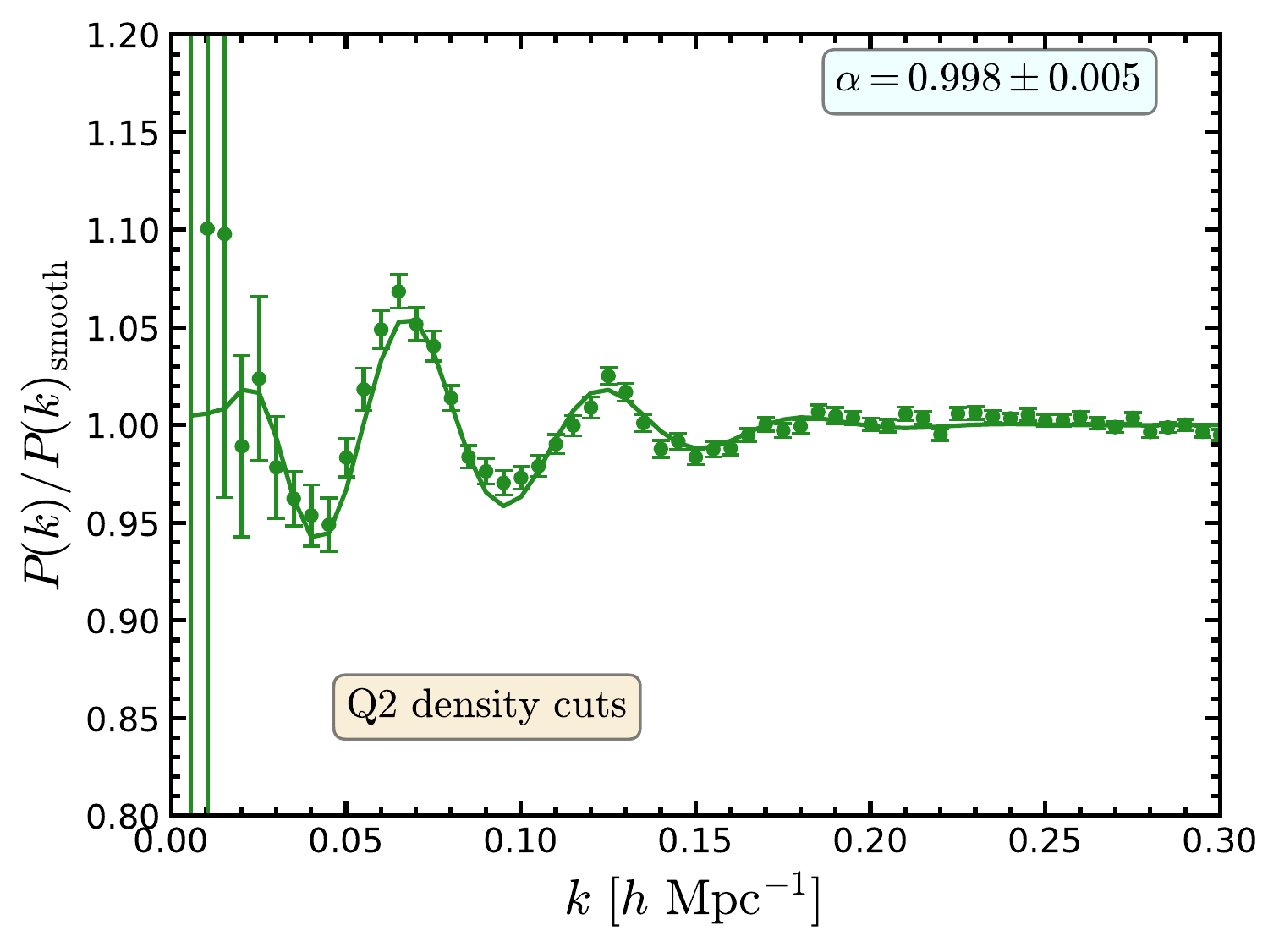}
    \includegraphics[width=0.33\textwidth]{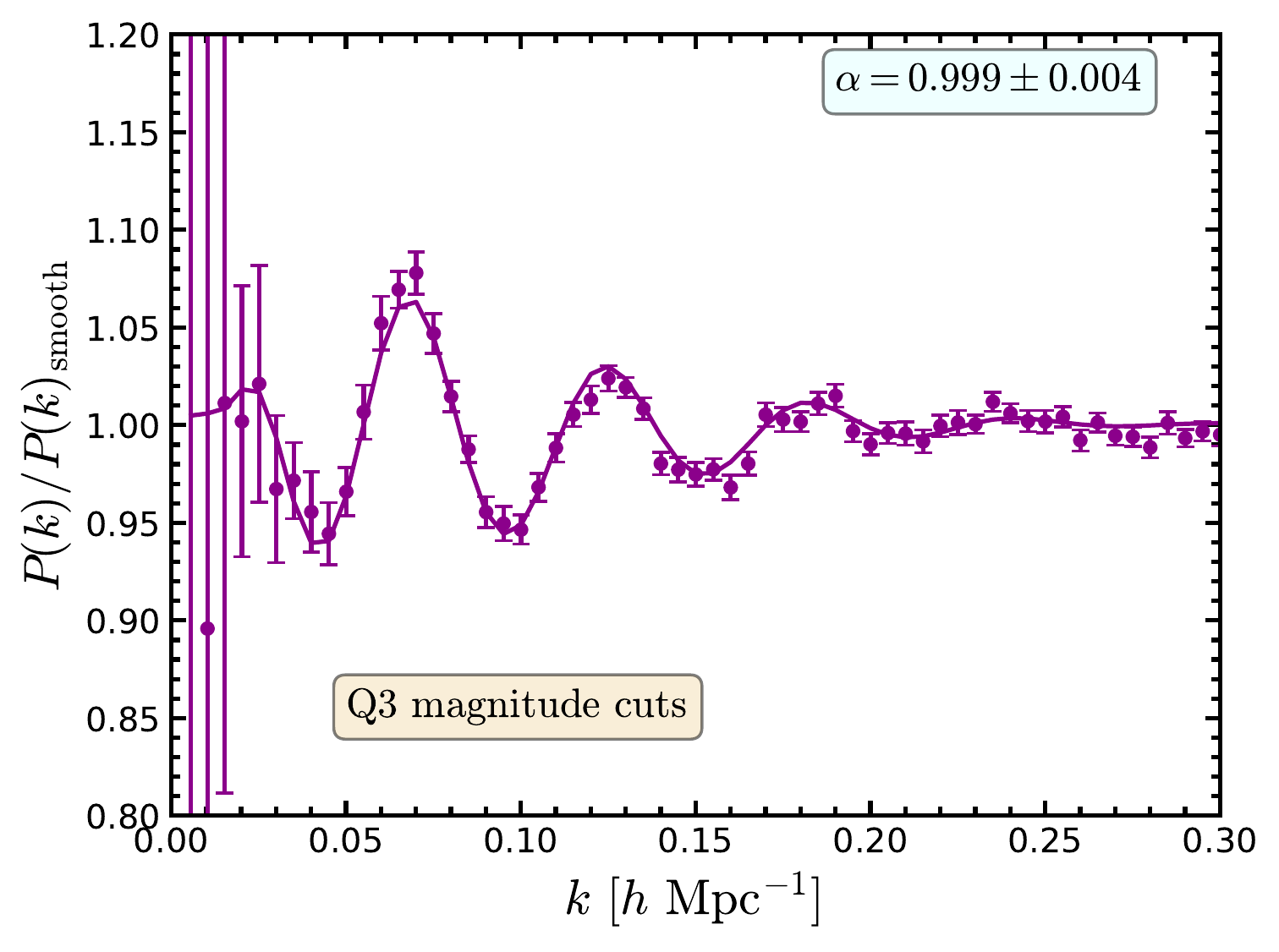}
    \includegraphics[width=0.33\textwidth]{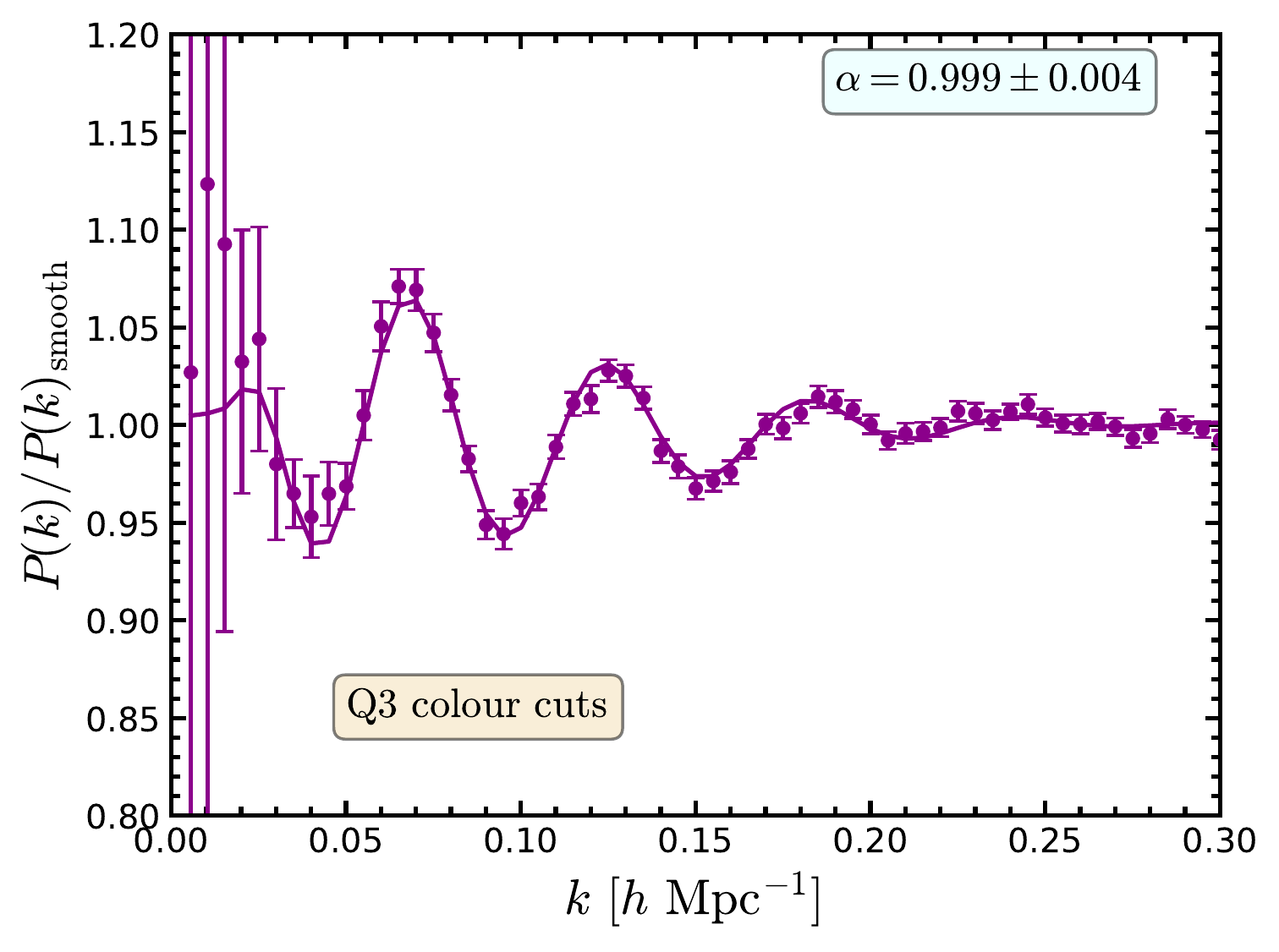}
    \includegraphics[width=0.33\textwidth]{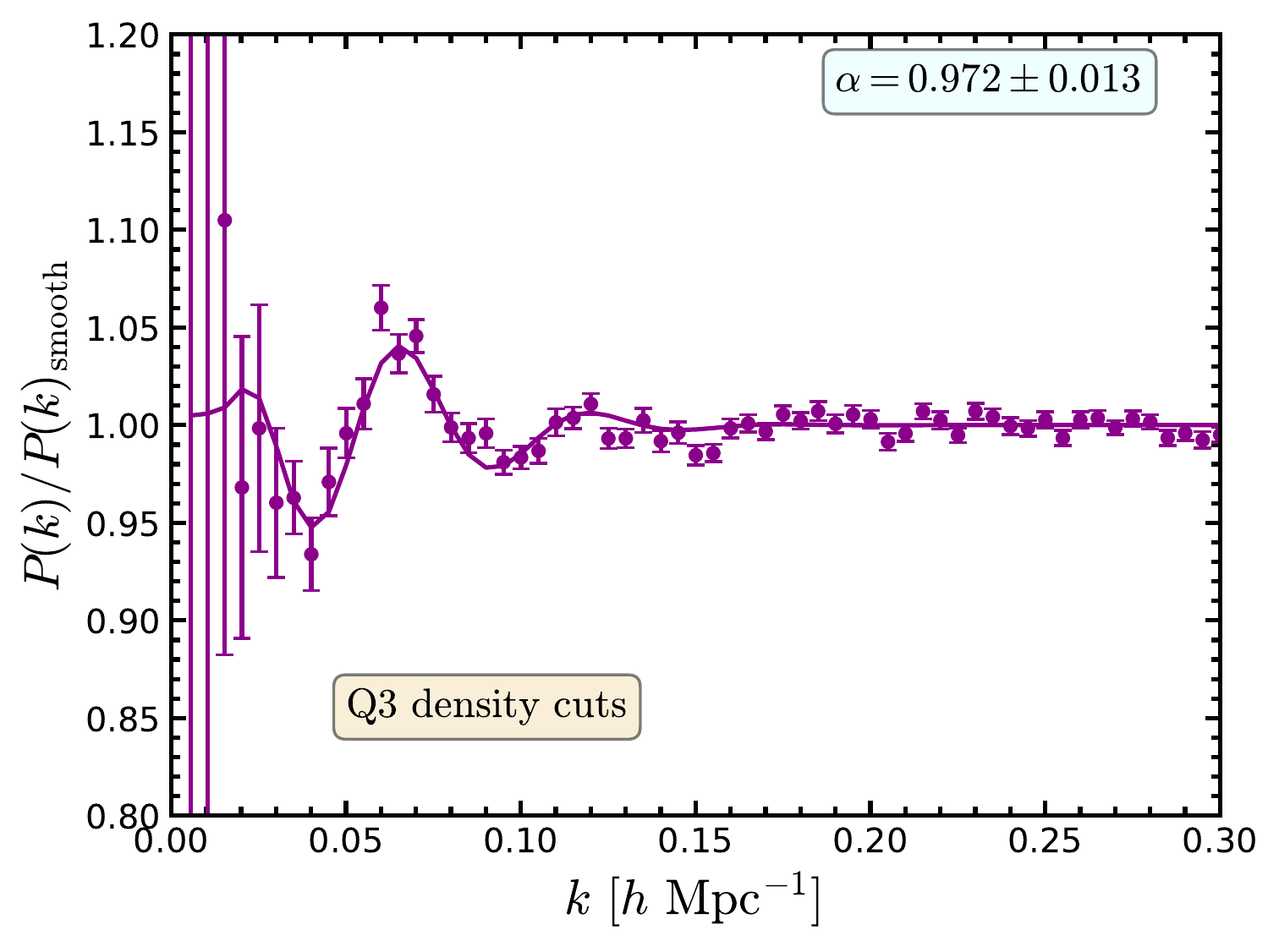}
    \includegraphics[width=0.33\textwidth]{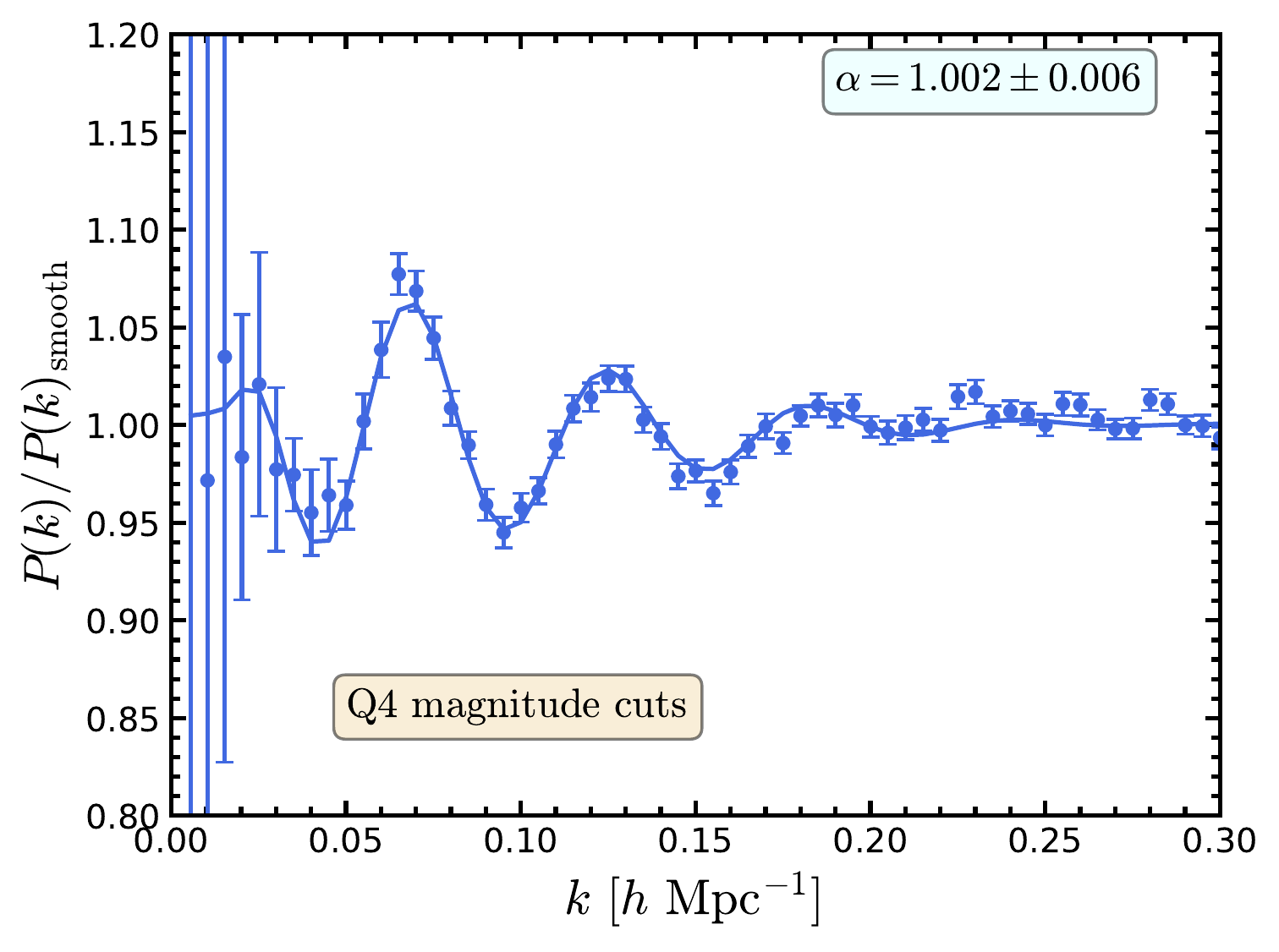}
    \includegraphics[width=0.33\textwidth]{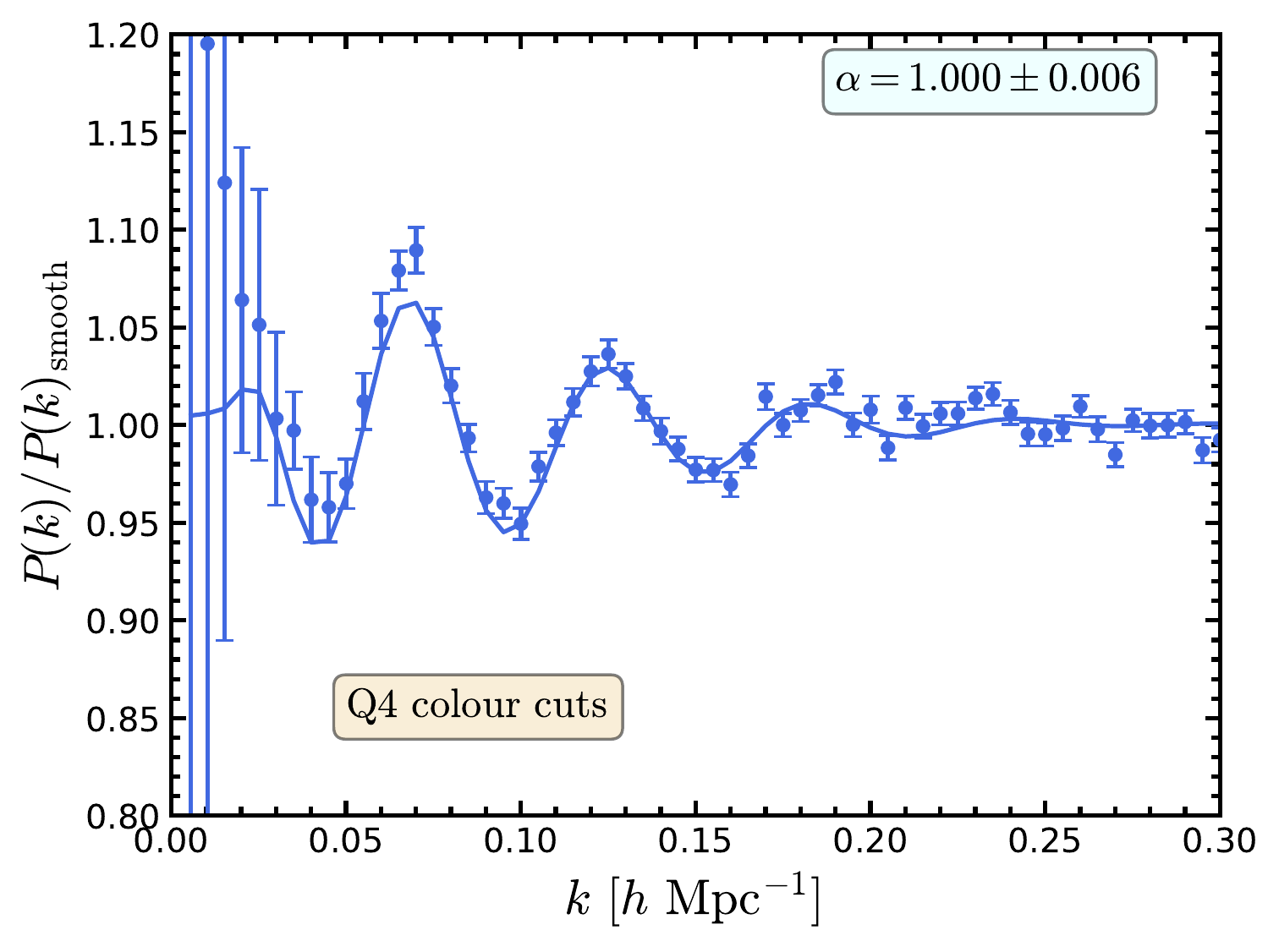}
    \includegraphics[width=0.33\textwidth]{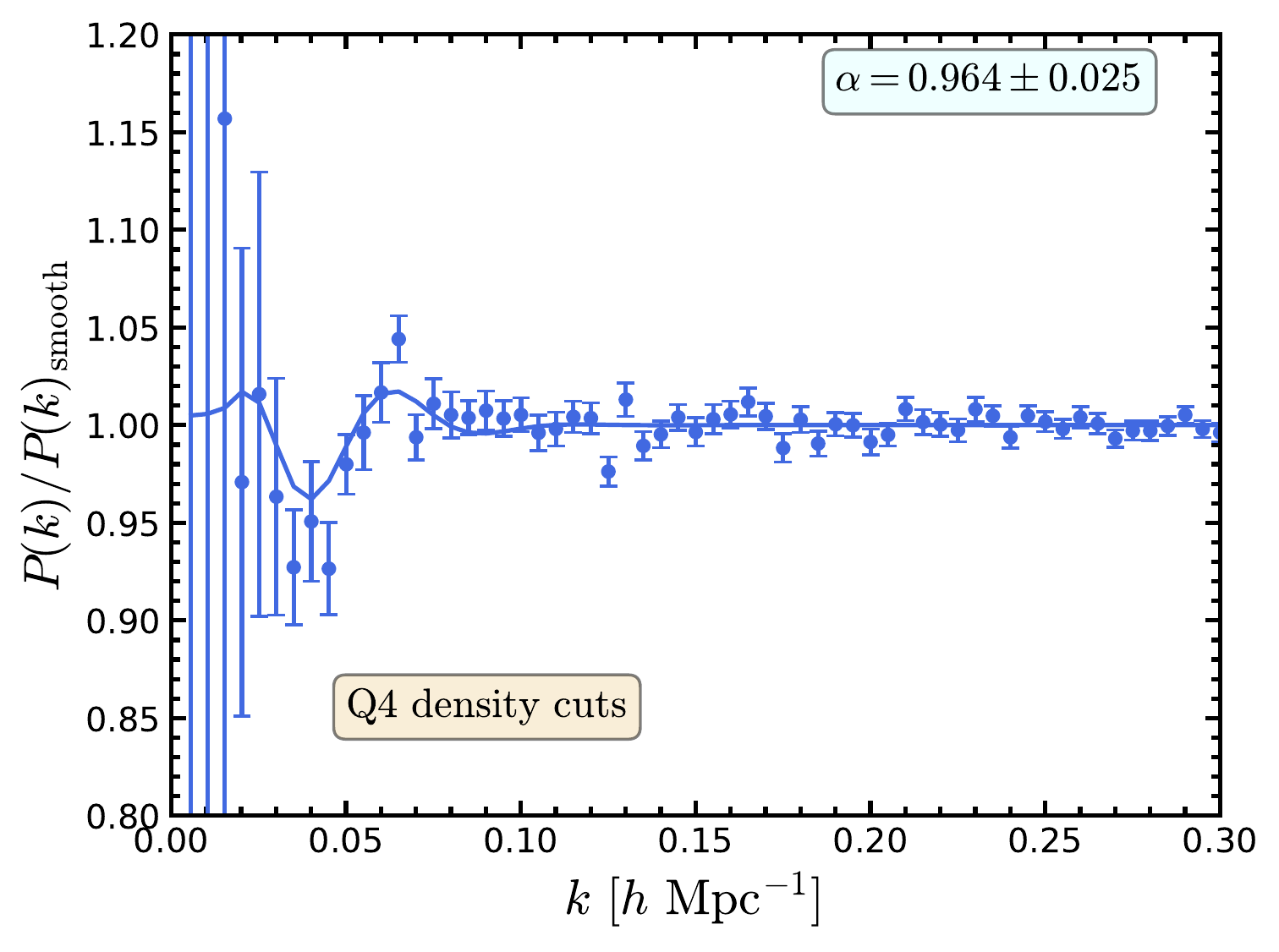}
    \vskip -.2cm
    \caption{The measured power spectrum, $P(k)$, (points with error bars) and the best-fitting model (solid curves) divided by the smooth (no-wiggle) power spectrum for magnitude ({\it left column}), colour ({\it middle column}) and density ({\it right column}) cuts. Each row shows a different subsample as specified in the bottom-left corner of each panel.  The strength of the BAO feature for each panel can be inferred from the uncertainties associated with the determination of the $\alpha$ dilation parameter (the maximum likelihood value and 68\% confidence interval of $\alpha$ are given in the top-right corner of each panel).} 
	\label{fig:fits_all}
\end{figure*}

\begin{figure}
    \centering
    \includegraphics[width=0.45\textwidth]{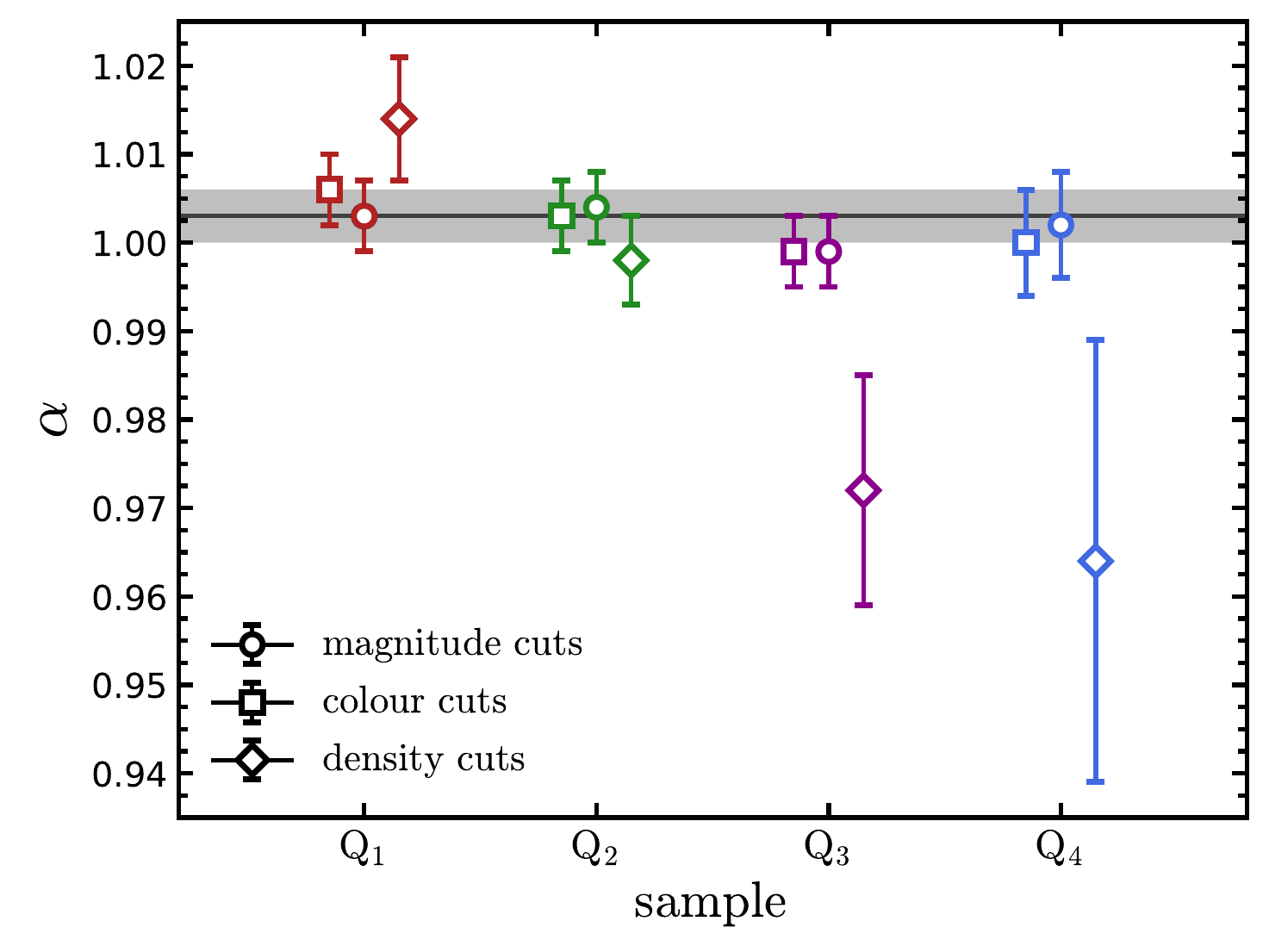}
    \caption{The maximum likelihood value and 68\% confidence interval associated with the determination of the BAO dilation parameter, $\alpha$. The horizontal solid line and the associated shaded region show the result for the full galaxy sample. The points with error bars show the results for the quartiles of the galaxy populations ranked according to: luminosity (circles), colour (squares) and density (diamonds). The Q1 subsamples corresponds to the brightest / reddest / densest galaxies, while the Q4 subsamples correspond to the faintest / bluest / least dense galaxies.}
	\label{fig:alpha_all}
\end{figure}

In the left panel of Fig.~\ref{fig:Pk_n1} we show the power spectrum measured from the full galaxy sample compared to the best-fitting  model. One can see that the model described by Eqs.~\eqref{eq:Pfit}-\eqref{eq:Odamp} provides a good match to the measurements from the mock catalogue.
The right panel of Fig.~\ref{fig:Pk_n1} displays the measured and best-fitting power spectra divided by the smooth component, $P_{\rm nw}(k)$, of the best-fitting model.
We recover an unbiased estimate of the BAO position, with $\alpha = 1.003 \pm 0.003$ (these figures correspond to the maximum likelihood value and $68$ per cent  confidence interval), that is consistent at the $1\sigma$-level with the expected value  of $\alpha=1$. Small differences in the value of $\alpha$ from unity are not necessarily worrying since they could indicate a small mismatch between the formulation of the power spectrum used to imprint the BAO feature onto the initial conditions of the simulation and the BAO templates used to extract this signature. The best-fitting model is characterised by a reduced chi-square value, $\chi^2/{\rm dof} = 1.15$, which indicates that our model gives a good description of the galaxy power spectrum. The quality of the fit is most clearly illustrated in the right panel of Fig.~\ref{fig:Pk_n1}, which also clearly highlights the BAO wiggles. In particular, we can see up to four maxima located at $k/(\hMpc) \approx 0.065$, $0.13$, $0.185$ and $0.24$. For the rest of the paper, we will compare the BAO position measured in the various galaxy subsamples against this reference value.

In Fig.~\ref{fig:fits_all} we show the results of fitting the BAO template, Eq.~\eqref{eq:Pfit}, to the various luminosity-, colour- and density-selected galaxy subsamples described in Section \ref{sec:selection}. To better highlight the quality of the fits and the changes in the BAO signature between the various subsamples, we show the power spectrum divided by the smooth component of the best-fitting model (see Eq.~\ref{eq:Psm}).

We find that all the magnitude- and colour-selected subsamples show the same BAO features, with little variation between the different subsamples. Considering the best-fitting $\alpha$ parameters, we find that most values are in good agreement, given the quoted interval, with the value measured for the full sample of $\alpha = 1.003 \pm 0.003$. There is potentially a very weak trend, such that fainter or bluer galaxies have slightly lower $\alpha$ values than their brighter or redder counterparts, but this trend is very small and we would need much larger galaxy samples to be able to confirm it. The only significant difference between the various quartiles is that the BAO signature is weaker for the Q4 samples, i.e. the faintest or bluest galaxies. This can be seen in the actual power spectrum (the fourth BAO wiggle is weaker for Q4 than in the other subsamples) and is best quantified by the uncertainty associated with the $\alpha$ measurement: the Q4 sample has an error on $\alpha$ of $0.6\%$ versus the errors of $0.4\%$ associated with the other quartiles. We also find that despite having four times fewer objects than the full sample, the $\alpha$ uncertainty ranges estimated for the Q1 to Q3 quartiles are only slightly larger than for the full sample ($0.4\%$ versus $0.3\%$). This means that the various quartiles are highly correlated and that increasing the sample size by a factor of four does not reduce the errors by half, as expected in the case of independently and Gaussianly distributed measurements.

The right column of Fig.~\ref{fig:fits_all} shows how the BAO signal varies for the four density-selected galaxy subsamples. Compared to the other two selection methods just discussed, we find that the density selection leads to much larger differences in the BAO signature compared to the full galaxy sample as well as between the different density quartiles. Firstly, we see that fewer BAO wiggles can be distinguished, for example the lowest density sample (${\rm Q}_4$) has one weak maximum, the ${\rm Q}_1$ and ${\rm Q}_3$ samples have two maxima, and ${\rm Q}_2$ has three maxima.
This is quite a striking difference, since in the full sample we clearly find four maxima (see right-hand panel in Fig.~\ref{fig:Pk_n1}). The smaller number of BAO wiggles for the highest density quartile, ${\rm Q}_1$, could be due to these galaxies residing in higher density regions where structure formation proceeds more rapidly and thus where non-linear effects, which dampen the BAO feature, are more pronounced. The result that the lowest density quartile, ${\rm Q}_4$, has only one BAO wiggle is more surprising, since, structure formation is somewhat delayed in lower density regions and thus more of the initial BAO signature should be preserved. However, we find that this is not the case.

The degradation in the BAO signal for the density-selected galaxy subsamples is best highlighted by comparing the uncertainties in determining $\alpha$ using the various quartiles. We find that the error is lowest for ${\rm Q}_2$ ($0.5\%$) and only slightly higher for ${\rm Q}_1$ ($0.7\%$), and increases dramatically for the lower density quartiles: $1.3\%$ and $2.5\%$  respectively for ${\rm Q}_3$ and ${\rm Q}_4$. Thus galaxies in intermediate-density environments (i.e. the ${\rm Q}_2$ quartile) are a better target to measure the BAO feature than those in the densest regions or least dense regions.  Furthermore, the uncertainty in determining $\alpha$ in the ${\rm Q}_2$ quartile is slightly larger than those associated with the luminosity- and colour-selected samples, indicating that selecting a galaxy subsample based on local density does {\it not} lead to a more precise BAO measurement than using colour or luminosity. In particular, the ${\rm Q}_1$ and ${\rm Q}_2$ density-quartiles have larger bias than the other luminosity- and colour-selected subsamples (see right column of Fig.~\ref{fig:Pk_samples}), implying that a sample with larger bias does not necessarily lead to a more precise determination of the BAO scale.

Another important result for the density-selected quartiles is that the $\alpha$ parameter systematically decreases with density. This is best illustrated in Fig.~\ref{fig:alpha_all}, which shows the maximum likelihood values and the 68\% confidence intervals on the determination of $\alpha$ for the various galaxy subsamples studied here. For the luminosity- and colour-selected quartiles the $\alpha$ value is approximately the same and in good agreement with the measurement obtained using the full galaxy sample. In contrast, the density-selected quartiles show a statistically significant trend that is in agreement with our expectations \citep[see e.g.][]{Sherwin:2012,Neyrinck:2016pfm}: the BAO peak is shifted to smaller scales (i.e. larger $\alpha$) for the densest quartile and to larger scales for the two least dense quartiles.

\subsection{Halo occupation distribution}\label{sec:hod}
\begin{figure*}
    \centering
    \includegraphics[width=0.458\textwidth]{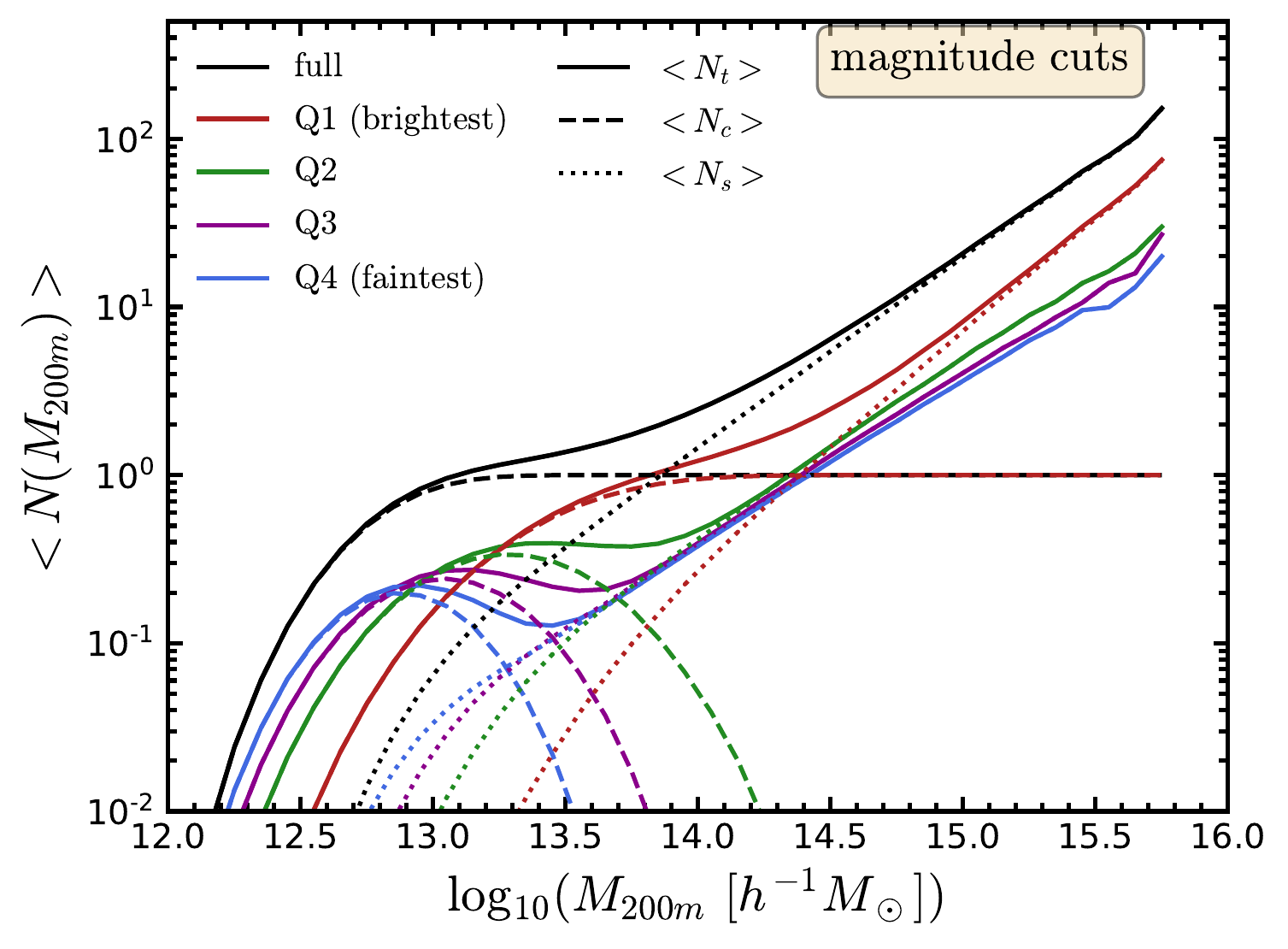}
    \includegraphics[width=0.45\textwidth]{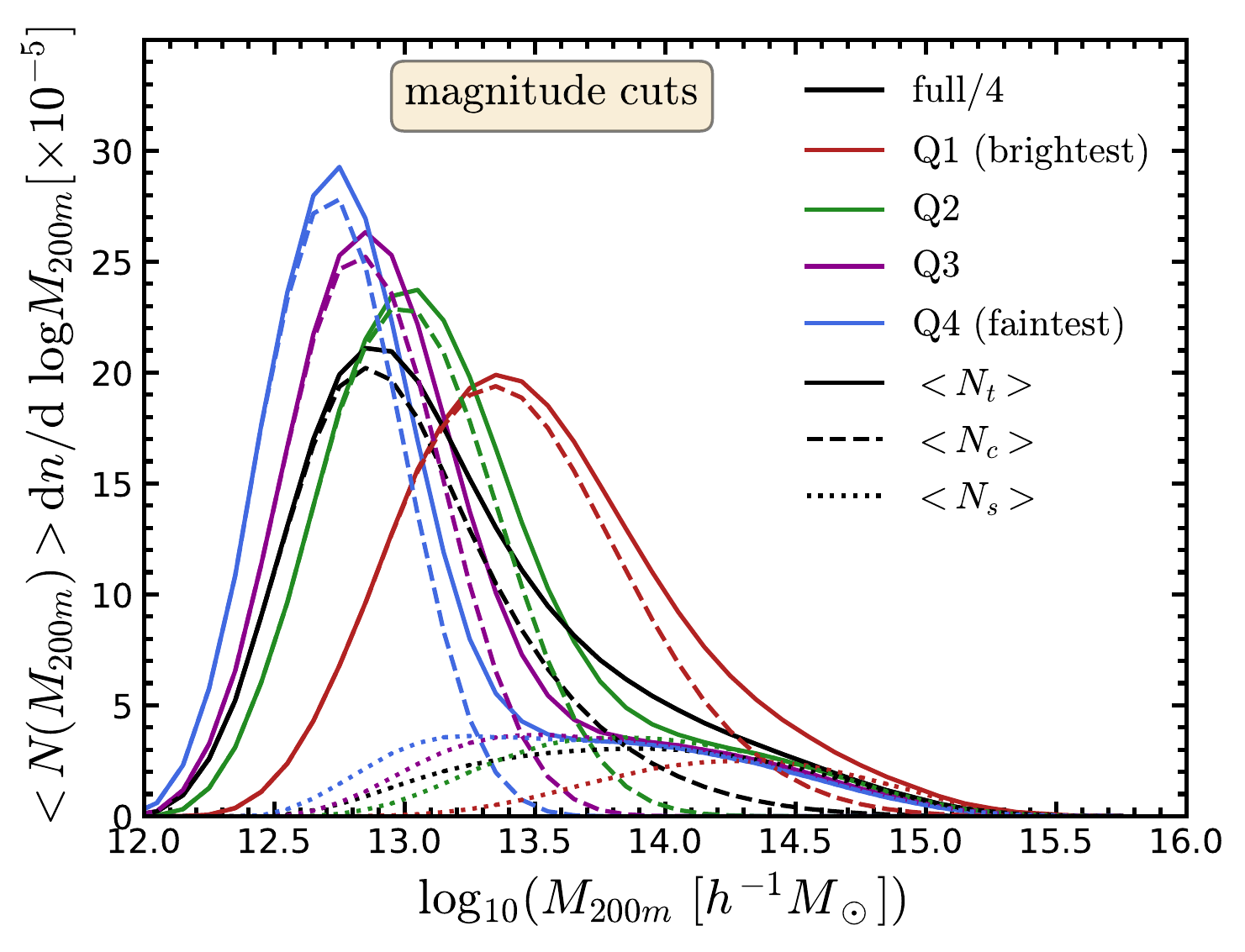}
    \includegraphics[width=0.458\textwidth]{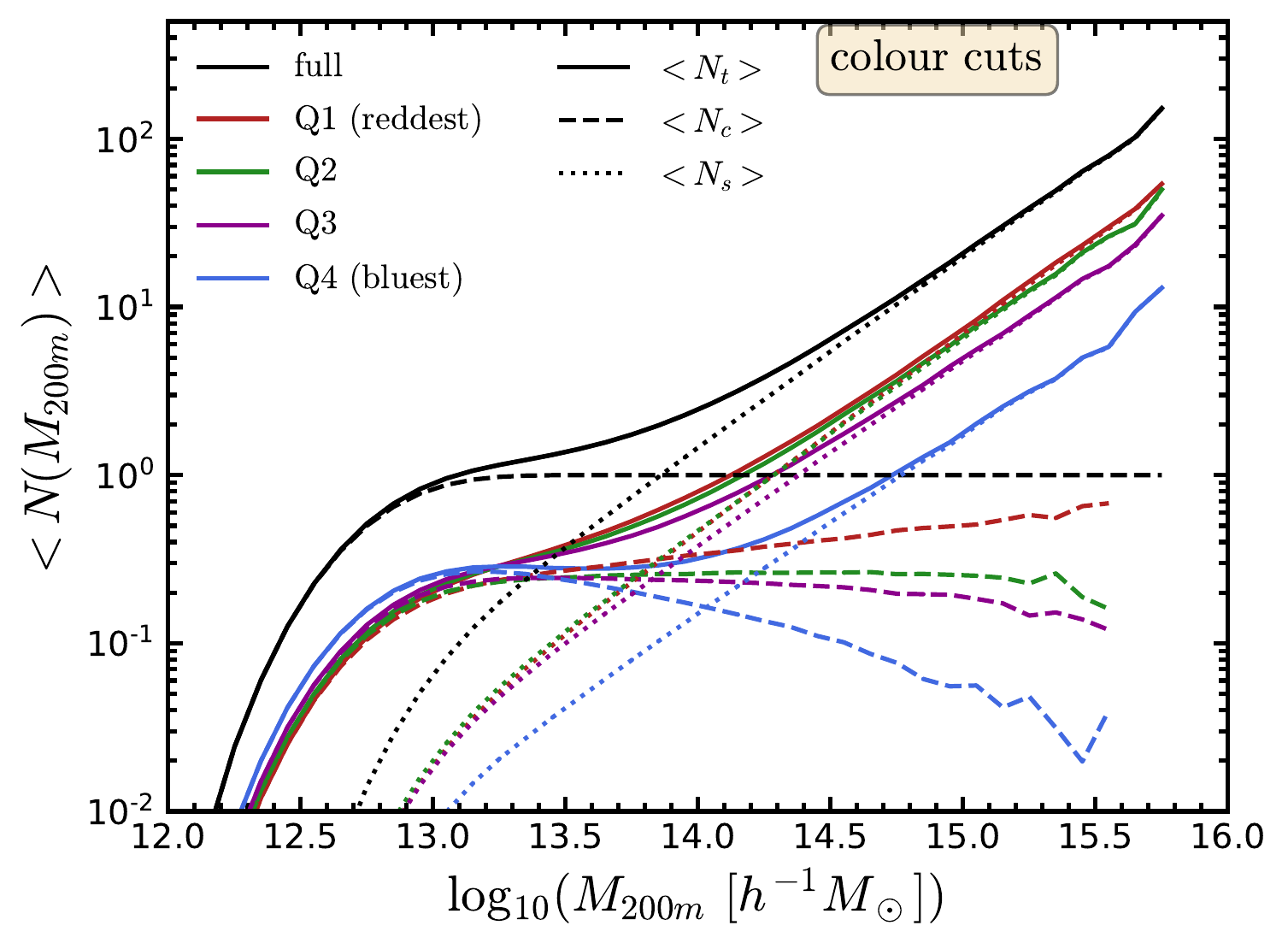}
    \includegraphics[width=0.45\textwidth]{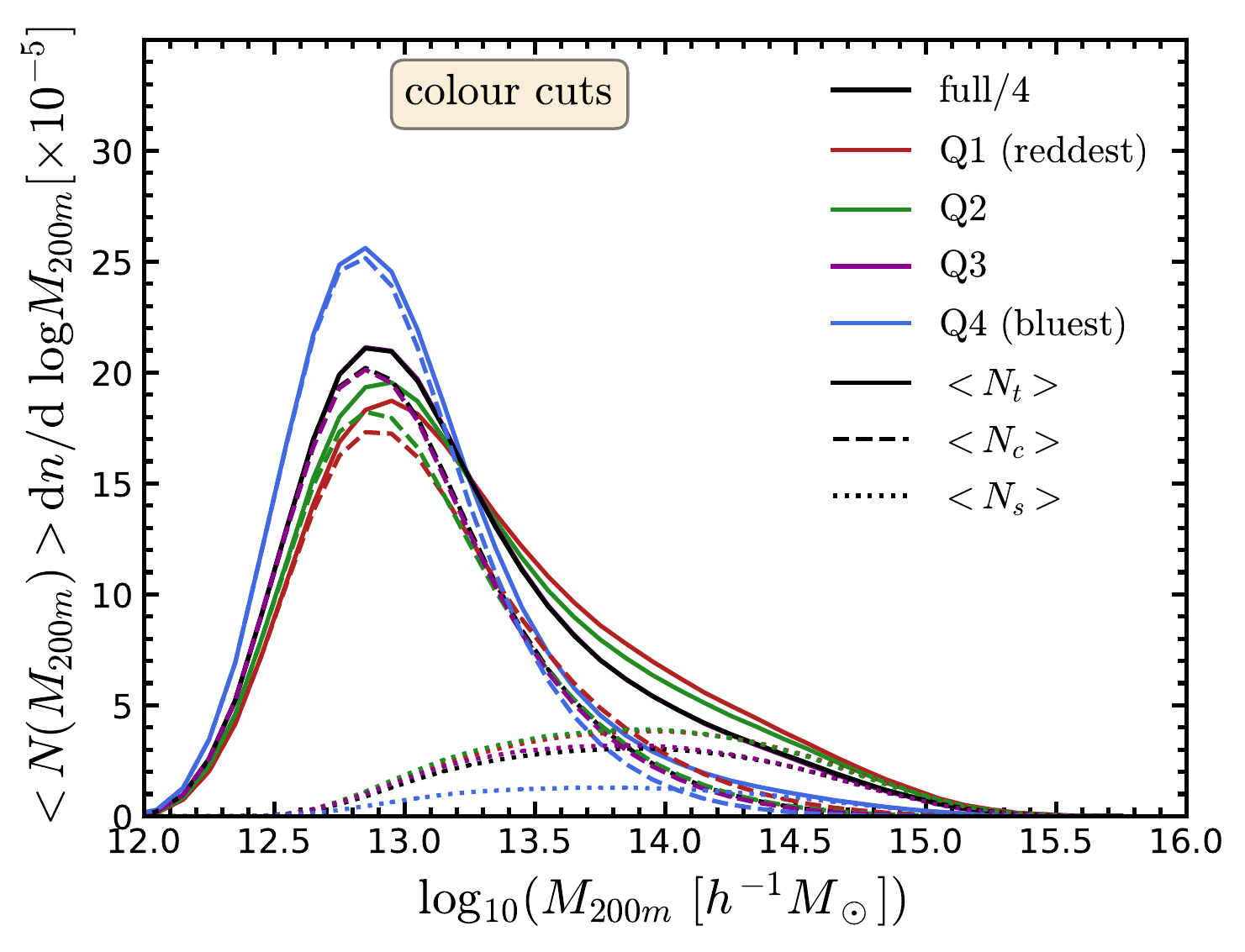}
    \includegraphics[width=0.458\textwidth]{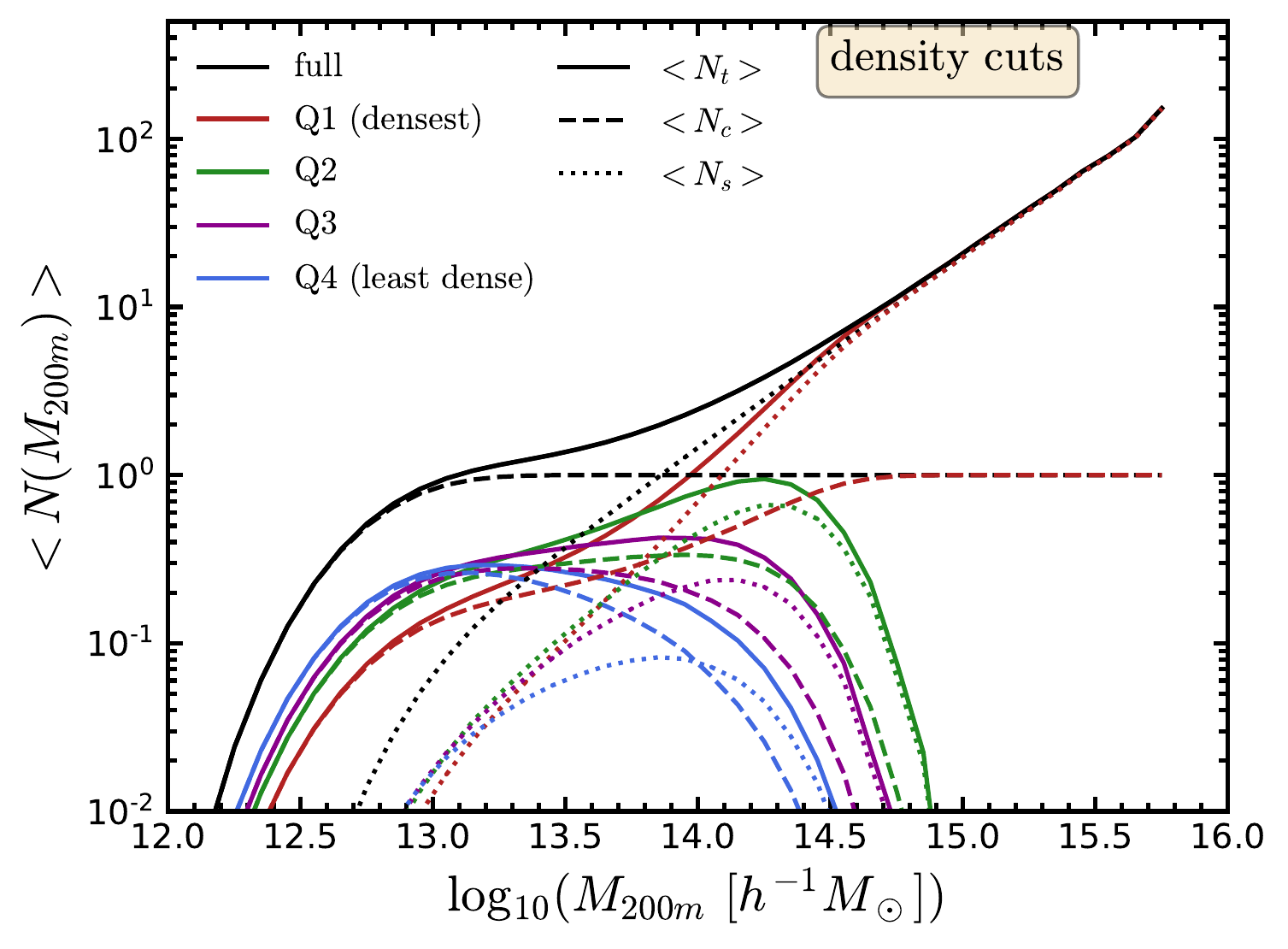}
    \includegraphics[width=0.45\textwidth]{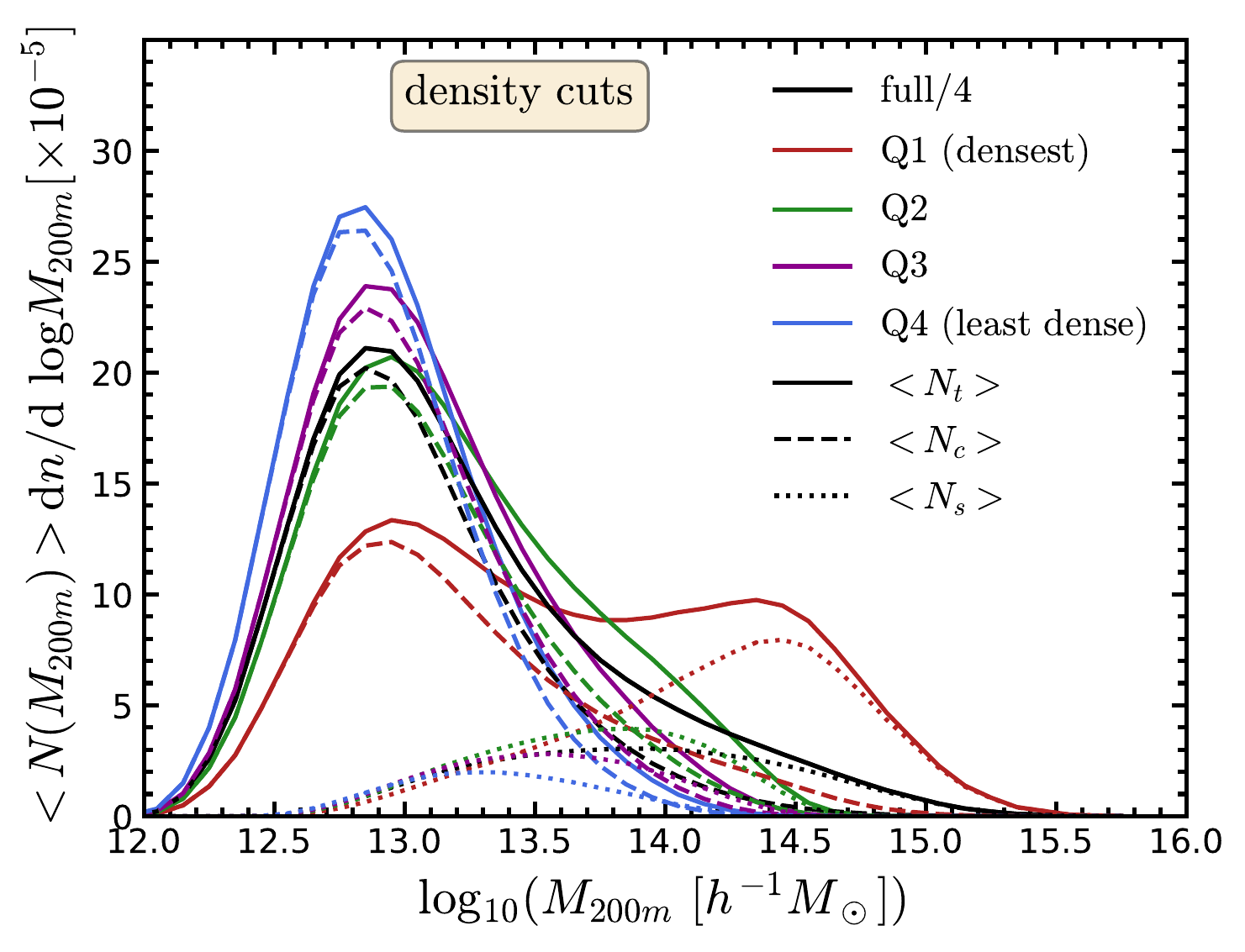}
    \caption{{\it Left column:} Halo occupation distribution for the galaxy quartiles, $Q_i$, selected according to: magnitude ({\it top panel}), colour ({\it middle panel}) and density ({\it bottom panel}). In each panel we show the HOD of the full sample (black lines) for comparison. The occupation of total, central and satellite galaxies are shown as solid, dashed and dotted lines, respectively, as specified in the legend. {\it Right column:} The number density of  central (dashed lines), satellite (dotted lines) and total (solid lines) galaxies for each selection and subsample obtained by multiplying the HOD by the differential halo mass function of the MXXL snapshot at $z=0.11$, in the case of the full sample we have divided the distribution by four for better visualisation. Different colours represent different subsamples: red lines (Q1), green lines (Q2), magenta lines (Q3) and blue lines (Q4).}
	\label{fig:HOD_samples}
\end{figure*}

To further investigate and understand the differences between the clustering results for different galaxy selections, we explore the halo occupation distribution in each subsample in the left column of Fig.~\ref{fig:HOD_samples}. In each case, we plot the contribution of central galaxies  (dashed lines), satellite galaxies (dotted lines) and the total mean number of galaxies per halo (solid lines), which is the sum of centrals and satellites. The HOD of the full sample is displayed by the black curves, while the contribution of different subsamples is shown by the red $({\rm Q}_1)$, green $({\rm Q}_2)$, magenta $({\rm Q}_3)$ and blue $({\rm Q}_4)$ curves in each panel.

The HOD of magnitude-selected galaxies is shown in the top left panel of Fig.~\ref{fig:HOD_samples}. We note that the HOD of the brightest galaxy quartile, ${\rm Q}_1$, is composed of galaxies that predominantly populate the most massive haloes, i.e., ${\rm Q}_1$ contains all the central galaxies of haloes with $\Mmean > 10^{14}\Msh$ and also the majority of the satellites found in these haloes. The fainter samples are composed of central galaxies in lower mass haloes and of satellite galaxies in high mass haloes. In particular, the mean number of satellites as a function of halo mass is roughly the same in the ${\rm Q}_2$, ${\rm Q}_3$ and ${\rm Q}_4$ quartiles, showing only a weak dependence on galaxy luminosity.

In the case of the HOD of galaxies ranked by colour (middle-left panel of Fig.~\ref{fig:HOD_samples}), we find a non-standard form for the mean number of central galaxies. For low halo masses, $\Mmean < 10^{13.4}\Msh$, there is a plateau at $\lla N_c \rra \sim 0.25$ for all quartiles. Interestingly, $\lla N_c \rra $ stays constant with increasing halo mass for the ${\rm Q}_2$ sample; for the ${\rm Q}_3$  and ${\rm Q}_4$ samples, the mean fraction of haloes with centrals declines for $\Mmean > 10^{13.5}\Msh$, and increases with halo mass for ${\rm Q}_1$. For satellites, we find similar $\lla N_s \rra$ values for the ${\rm Q}_1$, ${\rm Q}_2$ and ${\rm Q}_3$ quartiles, with only a weak trend with galaxy colour. In contrast, the bluest quartile contains a significantly lower mean satellite number for a given host halo mass.
We note that the HOD of the ${\rm Q}_2$ quartile has the same shape as the full sample but with mean values that are four times smaller; this might explain why this the BAO features measured for this sample best resemble those of the full galaxy population. In contrast, the ${\rm Q}_4$ quartile preferentially contains galaxies in low mass haloes (see middle-right panel of Fig.~\ref{fig:HOD_samples}), and has the weakest BAO signature.

The bottom-left panel of Fig.~\ref{fig:HOD_samples} shows the HOD of density-selected galaxy quartiles.
We see that the densest subsample (${\rm Q}_1$) contains almost {\it all} the satellite galaxies, as well as all the central galaxies that live in haloes more massive than $\log_{10}(\Mmean/\Msh) = 14.6$. 
Thus, a large fraction of ${\rm Q}_1$ galaxies are in clusters and other highly overdense regions, whose gravity pulls in the surrounding matter, which  explains why the BAO peak is shifted towards smaller scales for this sample.
We see that the ${\rm Q}_2$ sample contains no galaxies (centrals and satellites) which reside in haloes of mass $\log_{10}(\Mmean/\Msh) > 14.9$, and the distribution peaks at a total mean occupation number of almost one at $\log_{10}(\Mmean/\Msh) = 14.4$. In this subsample, galaxies are selected from intermediate-density regions, explaining the lack of galaxies in clusters. The ${\rm Q}_3$ and ${\rm Q}_4$ subsamples contain galaxies that populate low-mass haloes $(\log_{10}(\Mmean/\Msh) \approx 12.5 - 13.5)$ and are dominated by central galaxies. In these cases we can see that the fraction of satellite galaxies is small. These low-mass haloes represent small density peaks in the dark matter distribution, and typically live in regions like filaments and voids; these samples display a weak BAO signal, and the position of the peak is shifted to larger scales (we can see from the lower-left panel of Fig.~\ref{fig:Pk_samples} and Fig.~\ref{fig:xi_dens_all} that the BAO signal in the ${\rm Q}_4$ subsample is hard to discern).

The right panels of Fig.~\ref{fig:HOD_samples} show the number density of galaxies (in units of $10^{-5} h^{3} {\rm Mpc}^{-3}$) as a function of their host halo mass for the three selections: magnitude (top panel), colour (middle panel) and density (bottom panel), in all panels we show the distribution of galaxies for the full sample divided by four for comparison. The results presented in these panels confirm our findings from the HOD of the different selections. As an example, in magnitude-selected galaxies we can observe a trend in their distribution (top panel of right column in Fig.~\ref{fig:HOD_samples}), i.e., we can find more of the faintest (brightest) galaxies in low-(high-)mass haloes. In the case of colour-selected galaxies, the distribution of galaxies remains almost unchanged for the ${\rm Q}_1$, ${\rm Q}_2$ and ${\rm Q}_3$ samples; the bluest sample (${\rm Q}_4$) predominantly populate haloes with mass $\log_{10}(\Mmean/\Msh) \approx 12.7$. The bottom-right panel of Fig.~\ref{fig:HOD_samples} shows the distribution of galaxies ranked by environment, we can see that galaxies from low-density to intermediate-density regions reside in low-mass haloes, while galaxies in the densest environments are found in high-mass haloes.
\section{Summary and conclusions}\label{sec:conc}
We have studied the clustering and the position of the BAO feature for subsamples of mock galaxies ranked by density (defined by the distance to $10^{\rm th}$ nearest neighbour), luminosity ($r$-band magnitude) and $\gr$ colour. 

We have used a magnitude-limited, $\Mr < -21.08$, mock catalogue at redshift $z=0.11$, obtained from the Millennium-XXL N-body simulation \citep{Smith:2017tzz}. This corresponds to a galaxy number density of $n = 1\times 10^{-3} \hMpcc$ that, given the large volume of the simulation, includes a total of 27 million galaxies. We split the full sample into four subsamples (${\rm Q}_1$, ${\rm Q}_2$, ${\rm Q}_3$ and ${\rm Q}_4$), defined in different ways (see below) with a corresponding number density of one quarter of the full sample $(n_Q = 2.5 \times 10^{-4}\hMpcc)$ by applying cuts according to the galaxy property of interest (see Sec.~\ref{sec:selection}). The subsamples are defined as follows, 1) magnitude: from brightest to faintest galaxies, 2) colour: from reddest to bluest galaxies and 3) density: from high- to low-density regions. We confirmed that the galaxy bias of each subsample is constant on linear scales, $k \lesssim 0.1 \hMpc$. We have measured the power spectrum of each subsample and fit it to an analytical BAO template to extract the position of the BAO peak through the dilation parameter, $\alpha$ (see Eq.~\ref{eq:Pfit}).

We have found that the best-fitting value of $\alpha$ for the full sample is $\alpha = 1.003 \pm 0.003$ and in each subsample the best sample to extract the BAO peak position is, 1) magnitude: ${\rm Q}_1$ with $\alpha = 1.003 \pm 0.004$, 2) colour: ${\rm Q}_2$ with $\alpha = 1.003 \pm 0.004$ and 3) density: ${\rm Q}_2$ with $\alpha = 0.998 \pm 0.005$. In general, all measurements for the luminosity- and colour-selected galaxy subsamples are in good agreement with the reference value of the full sample. However, 
for density-selected galaxies, the ${\rm Q}_1$, ${\rm Q}_3$ and ${\rm Q}_4$ quartiles recover poorly the position of the peak and are characterised by large uncertainties in the recovered BAO scale.

We have studied the HOD of each subsample to understand what are the host haloes corresponding to various galaxy selections and how this affects the BAO signal measurement. The luminosity- and colour-selected samples have broadly similar HODs, with the most important differences being: i) the brightest quartile consists of mostly galaxies in the most massive haloes, and ii) the bluest quartile contains few galaxies in high mass haloes, with most objects residing in lower mass hosts. The density-selected quartiles show the largest difference in HOD distributions: the densest quartile contains all the central and satellite galaxies of high-mass haloes, while the lowest density quartile consists of galaxies which are predominantly in low-mass haloes.

Our main results can be summarised as follows,
\begin{itemize}
    \item The magnitude- and colour-selected samples have unbiased BAO signatures, i.e. the $\alpha$ dilation parameter is consistent with that of the full galaxy population, and the uncertainties with which the BAO peak can be measured are roughly the same for all the subsamples.
    The only exceptions are the faintest or the bluest quartiles, which have a $\sim 50$ per cent times larger error on $\alpha$ than the other subsamples. Note that for an equal number density of tracers there is a small increase in the precision of the BAO measurement if we were to select only the reddest galaxies, but the effect is minor.
    \item The density selected samples show several interesting effects. Firstly, the recovery of $\alpha$ is biased between the different quartiles: densest galaxies have $\alpha > 1$, while the lowest density ones have $\alpha < 1$. Secondly, the $\alpha$ uncertainties are lowest for the ${\rm Q}_1$ and $\rm{Q}_2$ quartiles, while the $\rm{Q}_3$ and $\rm{Q}_4$ samples provide much poorer BAO constraints.
    \item Selecting galaxies by density does not improve the BAO measurements compared to a similar number density sample selected by either magnitude or colour. 
    \item However, selecting galaxies by density shows the systematic shift in the BAO position expected for galaxies in overdense and underdense regions, as discussed in \cite{Neyrinck:2016pfm}. High density peaks lead to a contraction of the peak to smaller radii (i.e. larger $\alpha$), while low density region show an expansion of the BAO feature to larger radii (i.e. smaller $\alpha$).
\end{itemize}

We have found that selecting galaxies by either luminosity or colour does not introduce any systematic biases in the BAO signal. Such effects may have been  expected since galaxies show both a luminosity and colour segregation depending on their host halo mass, with brighter or redder galaxies preferentially populating the more massive haloes. The most massive haloes are mainly found in higher density regions, and thus potentially could be characterised by a contraction of the BAO peak at their position. If such a contraction exist, its size would be below the current uncertainties of this study, in which we have determined the BAO dilation parameter, $\alpha$, with a precision $\lesssim 0.4\%$. 

Our results are derived in the context of a HOD mock catalogue \citep{Smith:2017tzz} in which galaxies are assigned magnitudes ($r$-band) and colours ($\gr{}$) such that they provide a reasonable match to the projected two-point correlation function as measured in the SDSS and GAMA surveys \citep{Zehavi2011,Farrow2015}. It remains to be seen if the same conclusions are valid when using more complex and more physically realistic methods to populate haloes with galaxies, such as hydrodynamic simulations or semi-analytic models of galaxy formation. Due to computational demands, the former are not yet at a level where Gigaparsec volumes needed for BAO studies can easily be simulated, however semi-analytic models \citep[e.g.][]{Henriques2015,Lacey2016,Lagos2018,Baugh2019} look more promising on short time scales. Such studies will be crucial to characterise any systematic shift in the BAO position resulting from selecting galaxy subsamples based on luminosity, colour, environment or emission lines.

\section*{Acknowledgements}
We acknowledge helpful conversations with Sergio Contreras, Idit Zehavi and Tianxiang Mao. 
CH-A acknowledges support from the Mexican National Council of Science and Technology (CONACyT) through grant No. 286513/438352. 
MC acknowledges support by the EU Horizon 2020 research and innovation programme under a Marie Sk{\l}odowska-Curie grant agreement 794474 (DancingGalaxies) and by the ERC Advanced Investigator grant, DMIDAS [GA 786910].
BL is supported by an ERC Starting Grant, ERC-StG-PUNCA-716532. 
We acknowledge support from STFC Consolidated Grants ST/P000541/1, ST/L00075X/1. 
This work used the DiRAC@Durham facility managed by the Institute for Computational Cosmology on behalf of the STFC DiRAC HPC Facility (\url{www.dirac.ac.uk}). The equipment was funded by BEIS capital funding via STFC capital grants ST/K00042X/1, ST/P002293/1, ST/R002371/1 and ST/S002502/1, Durham University and STFC operations grant ST/R000832/1. DiRAC is part of the National e-Infrastructure.




\bibliographystyle{mnras}
\bibliography{ref} 

\begin{thebibliography}{}
\makeatletter
\relax
\def\mn@urlcharsother{\let\do\@makeother \do\$\do\&\do\#\do\^\do\_\do\%\do\~}
\def\mn@doi{\begingroup\mn@urlcharsother \@ifnextchar [ {\mn@doi@}
  {\mn@doi@[]}}
\def\mn@doi@[#1]#2{\def\@tempa{#1}\ifx\@tempa\@empty \href
  {http://dx.doi.org/#2} {doi:#2}\else \href {http://dx.doi.org/#2} {#1}\fi
  \endgroup}
\def\mn@eprint#1#2{\mn@eprint@#1:#2::\@nil}
\def\mn@eprint@arXiv#1{\href {http://arxiv.org/abs/#1} {{\tt arXiv:#1}}}
\def\mn@eprint@dblp#1{\href {http://dblp.uni-trier.de/rec/bibtex/#1.xml}
  {dblp:#1}}
\def\mn@eprint@#1:#2:#3:#4\@nil{\def\@tempa {#1}\def\@tempb {#2}\def\@tempc
  {#3}\ifx \@tempc \@empty \let \@tempc \@tempb \let \@tempb \@tempa \fi \ifx
  \@tempb \@empty \def\@tempb {arXiv}\fi \@ifundefined
  {mn@eprint@\@tempb}{\@tempb:\@tempc}{\expandafter \expandafter \csname
  mn@eprint@\@tempb\endcsname \expandafter{\@tempc}}}

\bibitem[\protect\citeauthoryear{{Achitouv} \& {Blake}}{{Achitouv} \&
  {Blake}}{2015}]{Achitouv:2015}
{Achitouv} I.,  {Blake} C.,  2015, \mn@doi [\prd] {10.1103/PhysRevD.92.083523},
  \href {https://ui.adsabs.harvard.edu/abs/2015PhRvD..92h3523A} {92, 083523}

\bibitem[\protect\citeauthoryear{{Alam} et~al.,}{{Alam}
  et~al.}{2017}]{Alam:2017hwk}
{Alam} S.,  et~al., 2017, \mn@doi [\mnras] {10.1093/mnras/stx721}, \href
  {http://adsabs.harvard.edu/abs/2017MNRAS.470.2617A} {470, 2617}

\bibitem[\protect\citeauthoryear{Amendola et~al.}{Amendola
  et~al.}{2013}]{Amendola:2012ys}
Amendola L.,  et~al., 2013, Living Rev.Rel., 16, 6

\bibitem[\protect\citeauthoryear{Anderson et~al.}{Anderson
  et~al.}{2014}]{Anderson:2013zyy}
Anderson L.,  et~al., 2014, \mn@doi [Mon. Not. Roy. Astron. Soc.]
  {10.1093/mnras/stu523}, 441, 24

\bibitem[\protect\citeauthoryear{{Angulo}, {Baugh}, {Frenk}  \&
  {Lacey}}{{Angulo} et~al.}{2008}]{Angulo:2008}
{Angulo} R.~E.,  {Baugh} C.~M.,  {Frenk} C.~S.,   {Lacey} C.~G.,  2008, \mn@doi
  [\mnras] {10.1111/j.1365-2966.2007.12587.x}, \href
  {https://ui.adsabs.harvard.edu/abs/2008MNRAS.383..755A} {383, 755}

\bibitem[\protect\citeauthoryear{Angulo, Springel, White, Jenkins, Baugh  \&
  Frenk}{Angulo et~al.}{2012}]{Angulo:2012ep}
Angulo R.~E.,  Springel V.,  White S. D.~M.,  Jenkins A.,  Baugh C.~M.,   Frenk
  C.~S.,  2012, \mn@doi [Mon. Not. Roy. Astron. Soc.]
  {10.1111/j.1365-2966.2012.21830.x}, 426, 2046

\bibitem[\protect\citeauthoryear{{Ata}, {Kitaura}  \& {M{\"u}ller}}{{Ata}
  et~al.}{2015}]{Ata2015}
{Ata} M.,  {Kitaura} F.-S.,   {M{\"u}ller} V.,  2015, \mn@doi [\mnras]
  {10.1093/mnras/stu2347}, \href
  {https://ui.adsabs.harvard.edu/abs/2015MNRAS.446.4250A} {446, 4250}

\bibitem[\protect\citeauthoryear{{Baugh} et~al.,}{{Baugh}
  et~al.}{2019}]{Baugh2019}
{Baugh} C.~M.,  et~al., 2019, \mn@doi [\mnras] {10.1093/mnras/sty3427}, \href
  {https://ui.adsabs.harvard.edu/abs/2019MNRAS.483.4922B} {483, 4922}

\bibitem[\protect\citeauthoryear{Beutler et~al.}{Beutler
  et~al.}{2017}]{Beutler:2016ixs}
Beutler F.,  et~al., 2017, \mn@doi [Mon. Not. Roy. Astron. Soc.]
  {10.1093/mnras/stw2373}, 464, 3409

\bibitem[\protect\citeauthoryear{{Birkin}, {Li}, {Cautun}  \& {Shi}}{{Birkin}
  et~al.}{2019}]{Birkin:2019}
{Birkin} J.,  {Li} B.,  {Cautun} M.,   {Shi} Y.,  2019, \mn@doi [\mnras]
  {10.1093/mnras/sty3365}, \href
  {https://ui.adsabs.harvard.edu/abs/2019MNRAS.483.5267B} {483, 5267}

\bibitem[\protect\citeauthoryear{Blake \& Glazebrook}{Blake \&
  Glazebrook}{2003}]{Blake:2003rh}
Blake C.,  Glazebrook K.,  2003, \mn@doi [Astrophys. J.] {10.1086/376983}, 594,
  665

\bibitem[\protect\citeauthoryear{Blas, Lesgourgues  \& Tram}{Blas
  et~al.}{2011}]{Blas:2011rf}
Blas D.,  Lesgourgues J.,   Tram T.,  2011, \mn@doi [JCAP]
  {10.1088/1475-7516/2011/07/034}, 1107, 034

\bibitem[\protect\citeauthoryear{{Cole} et~al.,}{{Cole}
  et~al.}{2005}]{Cole:2005}
{Cole} S.,  et~al., 2005, \mn@doi [\mnras] {10.1111/j.1365-2966.2005.09318.x},
  \href {https://ui.adsabs.harvard.edu/abs/2005MNRAS.362..505C} {362, 505}

\bibitem[\protect\citeauthoryear{{DESI Collaboration} et~al.,}{{DESI
  Collaboration} et~al.}{2016}]{DESI:2016}
{DESI Collaboration} et~al., 2016, arXiv e-prints, \href
  {https://ui.adsabs.harvard.edu/abs/2016arXiv161100036D} {p. arXiv:1611.00036}

\bibitem[\protect\citeauthoryear{{Drinkwater} et~al.,}{{Drinkwater}
  et~al.}{2010}]{Drinkwater:2010}
{Drinkwater} M.~J.,  et~al., 2010, \mn@doi [\mnras]
  {10.1111/j.1365-2966.2009.15754.x}, \href
  {https://ui.adsabs.harvard.edu/abs/2010MNRAS.401.1429D} {401, 1429}

\bibitem[\protect\citeauthoryear{{Efstathiou}, {Kaiser}, {Saunders},
  {Lawrence}, {Rowan-Robinson}, {Ellis}  \& {Frenk}}{{Efstathiou}
  et~al.}{1990}]{Esfathiou:1990}
{Efstathiou} G.,  {Kaiser} N.,  {Saunders} W.,  {Lawrence} A.,
  {Rowan-Robinson} M.,  {Ellis} R.~S.,   {Frenk} C.~S.,  1990, \mnras, \href
  {https://ui.adsabs.harvard.edu/abs/1990MNRAS.247P..10E} {247, 10P}

\bibitem[\protect\citeauthoryear{Eisenstein \& Hu}{Eisenstein \&
  Hu}{1998}]{Eisenstein:1997ik}
Eisenstein D.~J.,  Hu W.,  1998, \mn@doi [Astrophys. J.] {10.1086/305424}, 496,
  605

\bibitem[\protect\citeauthoryear{{Eisenstein} et~al.,}{{Eisenstein}
  et~al.}{2001}]{Eisenstein:2001}
{Eisenstein} D.~J.,  et~al., 2001, \mn@doi [\aj] {10.1086/323717}, \href
  {https://ui.adsabs.harvard.edu/abs/2001AJ....122.2267E} {122, 2267}

\bibitem[\protect\citeauthoryear{Eisenstein et~al.}{Eisenstein
  et~al.}{2005}]{Eisenstein:2005su}
Eisenstein D.~J.,  et~al., 2005, \mn@doi [Astrophys.J.] {10.1086/466512}, 633,
  560

\bibitem[\protect\citeauthoryear{{Eisenstein}, {Seo}  \& {White}}{{Eisenstein}
  et~al.}{2007a}]{Eisenstein2007}
{Eisenstein} D.~J.,  {Seo} H.-J.,   {White} M.,  2007a, \mn@doi [\apj]
  {10.1086/518755}, \href
  {https://ui.adsabs.harvard.edu/abs/2007ApJ...664..660E} {664, 660}

\bibitem[\protect\citeauthoryear{{Eisenstein}, {Seo}, {Sirko}  \&
  {Spergel}}{{Eisenstein} et~al.}{2007b}]{Eisenstein2007ApJ...664..675E}
{Eisenstein} D.~J.,  {Seo} H.-J.,  {Sirko} E.,   {Spergel} D.~N.,  2007b,
  \mn@doi [\apj] {10.1086/518712}, \href
  {https://ui.adsabs.harvard.edu/abs/2007ApJ...664..675E} {664, 675}

\bibitem[\protect\citeauthoryear{{Farrow} et~al.,}{{Farrow}
  et~al.}{2015}]{Farrow2015}
{Farrow} D.~J.,  et~al., 2015, \mn@doi [\mnras] {10.1093/mnras/stv2075}, \href
  {https://ui.adsabs.harvard.edu/abs/2015MNRAS.454.2120F} {454, 2120}

\bibitem[\protect\citeauthoryear{{Feldman}, {Kaiser}  \& {Peacock}}{{Feldman}
  et~al.}{1994}]{Feldman:1994}
{Feldman} H.~A.,  {Kaiser} N.,   {Peacock} J.~A.,  1994, \mn@doi [\apj]
  {10.1086/174036}, \href
  {https://ui.adsabs.harvard.edu/abs/1994ApJ...426...23F} {426, 23}

\bibitem[\protect\citeauthoryear{{Foreman-Mackey}, {Hogg}, {Lang}  \&
  {Goodman}}{{Foreman-Mackey} et~al.}{2013}]{emcee:2013}
{Foreman-Mackey} D.,  {Hogg} D.~W.,  {Lang} D.,   {Goodman} J.,  2013, \mn@doi
  [\pasp] {10.1086/670067}, \href
  {https://ui.adsabs.harvard.edu/abs/2013PASP..125..306F} {125, 306}

\bibitem[\protect\citeauthoryear{{Hada} \& {Eisenstein}}{{Hada} \&
  {Eisenstein}}{2018}]{Hada:2018}
{Hada} R.,  {Eisenstein} D.~J.,  2018, \mn@doi [\mnras]
  {10.1093/mnras/sty1203}, \href
  {https://ui.adsabs.harvard.edu/abs/2018MNRAS.478.1866H} {478, 1866}

\bibitem[\protect\citeauthoryear{Hand, Feng, Beutler, Li, Modi, Seljak  \&
  Slepian}{Hand et~al.}{2018}]{Hand:2017pqn}
Hand N.,  Feng Y.,  Beutler F.,  Li Y.,  Modi C.,  Seljak U.,   Slepian Z.,
  2018, \mn@doi [Astron. J.] {10.3847/1538-3881/aadae0}, 156, 160

\bibitem[\protect\citeauthoryear{{Henriques}, {White}, {Thomas}, {Angulo},
  {Guo}, {Lemson}, {Springel}  \& {Overzier}}{{Henriques}
  et~al.}{2015}]{Henriques2015}
{Henriques} B. M.~B.,  {White} S. D.~M.,  {Thomas} P.~A.,  {Angulo} R.,  {Guo}
  Q.,  {Lemson} G.,  {Springel} V.,   {Overzier} R.,  2015, \mn@doi [\mnras]
  {10.1093/mnras/stv705}, \href
  {https://ui.adsabs.harvard.edu/abs/2015MNRAS.451.2663H} {451, 2663}

\bibitem[\protect\citeauthoryear{{Jasche} \& {Lavaux}}{{Jasche} \&
  {Lavaux}}{2019}]{Jasche2019}
{Jasche} J.,  {Lavaux} G.,  2019, \mn@doi [\aap] {10.1051/0004-6361/201833710},
  \href {https://ui.adsabs.harvard.edu/abs/2019A&A...625A..64J} {625, A64}

\bibitem[\protect\citeauthoryear{{Kaiser}}{{Kaiser}}{1986}]{Kaiser:1986}
{Kaiser} N.,  1986, \mn@doi [\mnras] {10.1093/mnras/219.4.785}, \href
  {https://ui.adsabs.harvard.edu/abs/1986MNRAS.219..785K} {219, 785}

\bibitem[\protect\citeauthoryear{{Lacey} et~al.,}{{Lacey}
  et~al.}{2016}]{Lacey2016}
{Lacey} C.~G.,  et~al., 2016, \mn@doi [\mnras] {10.1093/mnras/stw1888}, \href
  {https://ui.adsabs.harvard.edu/abs/2016MNRAS.462.3854L} {462, 3854}

\bibitem[\protect\citeauthoryear{{Lagos}, {Tobar}, {Robotham}, {Obreschkow},
  {Mitchell}, {Power}  \& {Elahi}}{{Lagos} et~al.}{2018}]{Lagos2018}
{Lagos} C. d.~P.,  {Tobar} R.~J.,  {Robotham} A. S.~G.,  {Obreschkow} D.,
  {Mitchell} P.~D.,  {Power} C.,   {Elahi} P.~J.,  2018, \mn@doi [\mnras]
  {10.1093/mnras/sty2440}, \href
  {https://ui.adsabs.harvard.edu/abs/2018MNRAS.481.3573L} {481, 3573}

\bibitem[\protect\citeauthoryear{Laureijs et~al.}{Laureijs
  et~al.}{2011}]{Laureijs:2011gra}
Laureijs R.,  et~al., 2011, preprint (\mn@eprint {arXiv} {1110.3193})

\bibitem[\protect\citeauthoryear{{Lesgourgues}}{{Lesgourgues}}{2011}]{Lesgourgues:2011re}
{Lesgourgues} J.,  2011, arXiv e-prints, \href
  {https://ui.adsabs.harvard.edu/abs/2011arXiv1104.2932L} {p. arXiv:1104.2932}

\bibitem[\protect\citeauthoryear{Linder}{Linder}{2003}]{Linder:2003ec}
Linder E.~V.,  2003, \mn@doi [Phys. Rev.] {10.1103/PhysRevD.68.083504}, D68,
  083504

\bibitem[\protect\citeauthoryear{{Loveday}, {Peterson}, {Efstathiou}  \&
  {Maddox}}{{Loveday} et~al.}{1992}]{Loveday:1992}
{Loveday} J.,  {Peterson} B.~A.,  {Efstathiou} G.,   {Maddox} S.~J.,  1992,
  \mn@doi [\apj] {10.1086/171284}, \href
  {https://ui.adsabs.harvard.edu/abs/1992ApJ...390..338L} {390, 338}

\bibitem[\protect\citeauthoryear{{McCullagh}, {Neyrinck}, {Szapudi}  \&
  {Szalay}}{{McCullagh} et~al.}{2013}]{McCullagh:2013}
{McCullagh} N.,  {Neyrinck} M.~C.,  {Szapudi} I.,   {Szalay} A.~S.,  2013,
  \mn@doi [\apjl] {10.1088/2041-8205/763/1/L14}, \href
  {https://ui.adsabs.harvard.edu/abs/2013ApJ...763L..14M} {763, L14}

\bibitem[\protect\citeauthoryear{Neyrinck, Szapudi, McCullagh, Szalay, Falck
  \& Wang}{Neyrinck et~al.}{2018}]{Neyrinck:2016pfm}
Neyrinck M.~C.,  Szapudi I.,  McCullagh N.,  Szalay A.,  Falck B.,   Wang J.,
  2018, \mn@doi [Mon. Not. Roy. Astron. Soc.] {10.1093/mnras/sty1074}, 478,
  2495

\bibitem[\protect\citeauthoryear{{Norberg}, {Baugh}, {Gazta{\~n}aga}  \&
  {Croton}}{{Norberg} et~al.}{2009}]{Norberg:2009}
{Norberg} P.,  {Baugh} C.~M.,  {Gazta{\~n}aga} E.,   {Croton} D.~J.,  2009,
  \mn@doi [\mnras] {10.1111/j.1365-2966.2009.14389.x}, \href
  {https://ui.adsabs.harvard.edu/abs/2009MNRAS.396...19N} {396, 19}

\bibitem[\protect\citeauthoryear{{Padmanabhan}, {Xu}, {Eisenstein}, {Scalzo},
  {Cuesta}, {Mehta}  \& {Kazin}}{{Padmanabhan} et~al.}{2012}]{Padmanabhan2012}
{Padmanabhan} N.,  {Xu} X.,  {Eisenstein} D.~J.,  {Scalzo} R.,  {Cuesta} A.~J.,
   {Mehta} K.~T.,   {Kazin} E.,  2012, \mn@doi [\mnras]
  {10.1111/j.1365-2966.2012.21888.x}, \href
  {https://ui.adsabs.harvard.edu/abs/2012MNRAS.427.2132P} {427, 2132}

\bibitem[\protect\citeauthoryear{{Planck Collaboration} et~al.,}{{Planck
  Collaboration} et~al.}{2016}]{Planck:2016}
{Planck Collaboration} et~al., 2016, \mn@doi [\aap]
  {10.1051/0004-6361/201525830}, \href
  {https://ui.adsabs.harvard.edu/abs/2016A&A...594A..13P} {594, A13}

\bibitem[\protect\citeauthoryear{Ross, Samushia, Howlett, Percival, Burden  \&
  Manera}{Ross et~al.}{2015}]{Ross:2014qpa}
Ross A.~J.,  Samushia L.,  Howlett C.,  Percival W.~J.,  Burden A.,   Manera
  M.,  2015, \mn@doi [Mon. Not. Roy. Astron. Soc.] {10.1093/mnras/stv154}, 449,
  835

\bibitem[\protect\citeauthoryear{Ross et~al.}{Ross et~al.}{2017}]{Ross:2016gvb}
Ross A.~J.,  et~al., 2017, \mn@doi [Mon. Not. Roy. Astron. Soc.]
  {10.1093/mnras/stw2372}, 464, 1168

\bibitem[\protect\citeauthoryear{Ruggeri \& Blake}{Ruggeri \&
  Blake}{2019}]{Ruggeri:2019kjl}
Ruggeri R.,  Blake C.,  2019, arXiv e-prints, \href
  {https://ui.adsabs.harvard.edu/abs/2019arXiv190913011R} {p. arXiv:1909.13011}

\bibitem[\protect\citeauthoryear{{Sherwin} \& {Zaldarriaga}}{{Sherwin} \&
  {Zaldarriaga}}{2012}]{Sherwin:2012}
{Sherwin} B.~D.,  {Zaldarriaga} M.,  2012, \mn@doi [\prd]
  {10.1103/PhysRevD.85.103523}, \href
  {https://ui.adsabs.harvard.edu/abs/2012PhRvD..85j3523S} {85, 103523}

\bibitem[\protect\citeauthoryear{{Shi}, {Cautun}  \& {Li}}{{Shi}
  et~al.}{2018}]{Shi:2018}
{Shi} Y.,  {Cautun} M.,   {Li} B.,  2018, \mn@doi [\prd]
  {10.1103/PhysRevD.97.023505}, \href
  {https://ui.adsabs.harvard.edu/abs/2018PhRvD..97b3505S} {97, 023505}

\bibitem[\protect\citeauthoryear{{Skibba} \& {Sheth}}{{Skibba} \&
  {Sheth}}{2009}]{Skibba:2009}
{Skibba} R.~A.,  {Sheth} R.~K.,  2009, \mn@doi [\mnras]
  {10.1111/j.1365-2966.2008.14007.x}, \href
  {https://ui.adsabs.harvard.edu/abs/2009MNRAS.392.1080S} {392, 1080}

\bibitem[\protect\citeauthoryear{{Skibba}, {Sheth}, {Connolly}  \&
  {Scranton}}{{Skibba} et~al.}{2006}]{Skibba:2006}
{Skibba} R.,  {Sheth} R.~K.,  {Connolly} A.~J.,   {Scranton} R.,  2006, \mn@doi
  [\mnras] {10.1111/j.1365-2966.2006.10196.x}, \href
  {https://ui.adsabs.harvard.edu/abs/2006MNRAS.369...68S} {369, 68}

\bibitem[\protect\citeauthoryear{Smith, Cole, Baugh, Zheng, Angulo, Norberg  \&
  Zehavi}{Smith et~al.}{2017}]{Smith:2017tzz}
Smith A.,  Cole S.,  Baugh C.,  Zheng Z.,  Angulo R.,  Norberg P.,   Zehavi I.,
   2017, \mn@doi [Mon. Not. Roy. Astron. Soc.] {10.1093/mnras/stx1432}, 470,
  4646

\bibitem[\protect\citeauthoryear{Spergel et~al.}{Spergel
  et~al.}{2003}]{Spergel:2003cb}
Spergel D.~N.,  et~al., 2003, \mn@doi [Astrophys. J. Suppl.] {10.1086/377226},
  148, 175

\bibitem[\protect\citeauthoryear{{Springel}, {White}, {Tormen}  \&
  {Kauffmann}}{{Springel} et~al.}{2001}]{Springel2001}
{Springel} V.,  {White} S. D.~M.,  {Tormen} G.,   {Kauffmann} G.,  2001,
  \mn@doi [\mnras] {10.1046/j.1365-8711.2001.04912.x}, \href
  {https://ui.adsabs.harvard.edu/abs/2001MNRAS.328..726S} {328, 726}

\bibitem[\protect\citeauthoryear{Springel et~al.}{Springel
  et~al.}{2005}]{Springel:2005nw}
Springel V.,  et~al., 2005, \mn@doi [Nature] {10.1038/nature03597}, 435, 629

\bibitem[\protect\citeauthoryear{Xu, Cuesta, Padmanabhan, Eisenstein  \&
  McBride}{Xu et~al.}{2013}]{Xu:2012fw}
Xu X.,  Cuesta A.~J.,  Padmanabhan N.,  Eisenstein D.~J.,   McBride C.~K.,
  2013, \mn@doi [Mon. Not. Roy. Astron. Soc.] {10.1093/mnras/stt379}, 431, 2834

\bibitem[\protect\citeauthoryear{{Zehavi} et~al.,}{{Zehavi}
  et~al.}{2011}]{Zehavi2011}
{Zehavi} I.,  et~al., 2011, \mn@doi [\apj] {10.1088/0004-637X/736/1/59}, \href
  {https://ui.adsabs.harvard.edu/abs/2011ApJ...736...59Z} {736, 59}

\bibitem[\protect\citeauthoryear{Zheng et~al.,}{Zheng
  et~al.}{2005}]{Zheng:2004id}
Zheng Z.,  et~al., 2005, \mn@doi [Astrophys. J.] {10.1086/466510}, 633, 791

\bibitem[\protect\citeauthoryear{{Zhu}, {Yu}, {Pen}, {Chen}  \& {Yu}}{{Zhu}
  et~al.}{2017}]{Zhu2017}
{Zhu} H.-M.,  {Yu} Y.,  {Pen} U.-L.,  {Chen} X.,   {Yu} H.-R.,  2017, \mn@doi
  [\prd] {10.1103/PhysRevD.96.123502}, \href
  {https://ui.adsabs.harvard.edu/abs/2017PhRvD..96l3502Z} {96, 123502}

\makeatother
\end{thebibliography}





\bsp	
\label{lastpage}
\end{document}